\documentclass[prd, reprint, amsmath, amssymb, nofootinbib, floatfix, aps, longbibliography, superscriptaddress]{revtex4-1}
\usepackage{graphicx}
\usepackage{dcolumn}
\usepackage{bm}
\usepackage[colorlinks=true,urlcolor=rossos,anchorcolor=black,citecolor=mygreen,filecolor=black,linkcolor=black,menucolor=blue,linktocpage=true]{hyperref} % should be commented out if the tex file will be compiled with latex in arXiv!!! (pdflatex is fine)
\newcommand{\apss}{Astrophysics and Space Science}
\usepackage{subfigure}
\usepackage[utf8x]{inputenc}

\usepackage{booktabs}
\usepackage{colortbl}
\usepackage{collcell}
\usepackage{enumerate}
\usepackage{float}
\usepackage{verbatim}
\usepackage{multirow}
\usepackage{xparse}
\usepackage{tabularx}

\graphicspath{{figs/}}

\usepackage{multirow}
\newcommand{\gag}{g_{a\gamma}}

\definecolor{rossos}{cmyk}{0,1,1,0.55}
\definecolor{mygreen}{rgb}{0.27, 0.64, 0.48}
\definecolor{mygray}{gray}{.95}

\begin{document}

% Use the \preprint command to pLace your local institutional report
%\preprint{}

\title{Optical circular polarization induced by axionlike particles in blazars}
%\title{Searching the axionlike Particles through the Blazar Optical Circular Polarization Measurements}


\author{Run-Min Yao}
\email{yaorunmin@ihep.ac.cn}
\affiliation{Key Laboratory of Particle Astrophysics, Institute of High Energy Physics, Chinese Academy of Sciences, Beijing, China}
\affiliation{School of Physical Sciences, University of Chinese Academy of Sciences, Beijing, China}

\author{Xiao-Jun Bi}
\email{bixj@ihep.ac.cn}
\affiliation{Key Laboratory of Particle Astrophysics, Institute of High Energy Physics, Chinese Academy of Sciences, Beijing, China}
\affiliation{School of Physical Sciences, University of Chinese Academy of Sciences, Beijing, China}

\author{Jin-Wei Wang}
\email{wangjinwei@sjtu.edu.cn (corresponding author)}
\affiliation{Tsung-Dao Lee Institute and School of Physics and Astronomy,
Shanghai Jiao Tong University, 800 Dongchuan Road, Shanghai 200240, China}
\affiliation{Scuola Internazionale Superiore di Studi Avanzati (SISSA), via Bonomea 265, 34136 Trieste, Italy}
\affiliation{INFN, Sezione di Trieste, via Valerio 2, 34127 Trieste, Italy}
\affiliation{Institute for Fundamental Physics of the Universe (IFPU), via Beirut 2, 34151 Trieste, Italy}

\author{Peng-Fei Yin}
\email{yinpf@ihep.ac.cn (corresponding author)}
\affiliation{Key Laboratory of Particle Astrophysics, Institute of High Energy Physics, Chinese Academy of Sciences, Beijing, China}

%\thanks{}

\date{\today}

\begin{abstract}
We propose that the interaction between the axionlike particles (ALPs) and photons can be a possible origin of optical circular polarization (CP) in blazars. Given that there is no definite detection of optical CP at $\sim0.1\%$ level, a rough limit on ALP-photon coupling can be obtained, specifically $\gag\cdot B_\mathrm{T0}\lesssim7.9\times10^{-12}~\mathrm{G\cdot GeV}^{-1}$ for $m_{a}\lesssim 10^{-13}~\mathrm{eV}$, depending on the magnetic field configuration of the blazar jet. Obviously, for the blazar models with a larger magnetic field strength, such as hadronic radiation models, this constraint could be more stringent. We also perform a dedicated analysis of the tentative observations of optical CP in two blazars, namely 3C 66A and OJ 287, and we find that these observations could be explained by the ALP-photon mixing with  $g_{a\gamma} \sim 10^{-11}~\mathrm{GeV}^{-1}$. As an outlook, our analysis can be improved by further research on the radiation models of blazars and high-precision joint measurements of optical CP and linear polarization.
\end{abstract}
%\keywords{}
\maketitle
\section{\label{sec:introduction}Introduction}
Axionlike particles (ALPs) are very light pseudoscalar particles that appear in many extensions of the Standard Model (SM) \cite{Witten:1984dg,Svrcek:2006yi,Conlon:2006tq,Conlon:2006ur,Arvanitaki:2009fg,Acharya:2010zx,Cicoli:2012sz}. These particles constitute a generalization of the quantum chromodynamics (QCD) axion, which is a pseudo-Goldstone boson arising from the Peccei-Quinn symmetry breaking mechanism and can be treated as an excellent solution to the strong \textit{CP}\footnote{The italic font \textit{CP} means charge-parity here. The readers should not confuse it with the normal abbreviation CP for circular polarization in this paper.} problem \cite{Peccei:1977hh,Peccei:1977ur,Weinberg:1977ma,Wilczek:1977pj,Cheng:1987gp}. Meanwhile, unlike the QCD axion, the coupling and mass of the ALP are, in principle, independent parameters, thus a much wider parameter space is spanned in the ALP scenario. ALPs are extremely appealing as they can also play important roles in cosmology and astrophysics \cite{Kolb:1990vq,Turner:1989vc,Sikivie:2006ni,Carroll:1998zi}. For instance, ALPs are good candidates for dark matter \cite{Marsh:2015xka,Preskill:1982cy,Abbott:1982af,Dine:1982ah}.

At low energies, ALPs could interact with the SM particles through the effective operators that are suppressed by some high energy
scales \cite{Graham:2015ouw}. For example, ALPs may interact with the electromagnetic sector through the Lagrangian
term $\mathcal{L} = \gag a\bm{E}\cdot\bm{B}$.
%While the axion mass is strictly related to its coupling with the standard model particles, ALPs possess a wide parameter space where the ALP mass $m_a$ could be independent on the coupling $\gag$.
This interaction can lead to two distinct effects in astrophysics, which provide promising ways to detect ALPs. One is the polarization rotation of photons as they propagate in a variable ALP background field, due to the change of the dispersion relation of the photons \cite{Fedderke:2019ajk,Schwarz:2020jjh,Ivanov:2018byi,DeRocco:2018jwe,Obata:2018vvr,Caputo:2019tms,Chen:2019fsq,Yuan:2020xui}. The other is the ALP-photon conversion in the external magnetic field \cite{Sikivie:1983ip,Raffelt:1987im}. The mixing occurs between the ALP and the photon polarization component that in parallel to the magnetic field. In particular, this process changes not only the amplitude but also the polarization of the photon \cite{Raffelt:1987im,Maiani:1986md}. %In the study of the ALP-photon mixing, the knowledge about the magnetic field where the ALP-photon system propagates is required. %Therefore, only fields that can be measured independently are considered.
Considering that the ALP-photon coupling is particularly weak, many studies focus on the conversion for high energy photons \cite{Csaki:2003ef,DeAngelis:2007dqd,Hooper:2007bq,Simet:2007sa,Sanchez-Conde:2009exi,Mirizzi:2009aj,Bassan:2010ya,DeAngelis:2011id,Galanti:2018upl,Dessert:2020lil} or the resonant conversion in strong magnetic fields \cite{Pshirkov:2007st,Hook:2018iia,Huang:2018lxq,Dessert:2019sgw,Darling:2020plz,Edwards:2020afl,Wang:2021wae,Wang:2021hfb,Prabhu:2021zve,Dessert:2022yqq}. Otherwise, the magnetic fields in large scales, e.g., the supercluster at scale of $\sim10~\mathrm{Mpc}$, are needed to achieve a significant conversion \cite{Jain:2002vx,Agarwal:2008ac,Payez:2011sh,Payez:2012vf,Tiwari:2016cps}. In particular, the ALP-photon mixing in extragalactic magnetic fields is proposed to explain the dimming of type Ia supernovae \cite{Csaki:2001yk,Grossman:2002by,Mirizzi:2006zy,Avgoustidis:2010ju,Liao:2015ccl,Buen-Abad:2020zbd}.

Many studies have addressed the ALP effects on the observed spectrum and polarization state of $\gamma$-rays from sources like blazars in astrophysical magnetic fields \cite{Csaki:2003ef,DeAngelis:2007dqd,Hooper:2007bq,Simet:2007sa,Sanchez-Conde:2009exi,Mirizzi:2009aj,Bassan:2010ya,DeAngelis:2011id,Galanti:2018upl,Meyer:2014epa,Tavecchio:2014yoa,Day:2018ckv,Galanti:2022iwb,Galanti:2022yxn,Galanti:2022tow}, including those in the blazar jet, galaxy cluster, intergalactic space, and Milky Way. However, the mixing effect is rarely discussed for low energy photons, especially within the source region of blazars.
%It is expected that the ALP-photon mixing would not induce a significant effect on the state of low energy photons under non-extreme conditions.
In fact, there are some advantages to study such effect. It is easier to measure the polarization of low energy photons with high precision \cite{Krawczynski:2015mud,10.1117/12.2275485}. Besides, with the multiwavelength observations of blazars, we have a better understanding of their properties, such as the radiation mechanism and the magnetic field structures \cite{Urry:1995mg,Angel:1980yg}, which in turn helps us to understand the initial polarization state of photons. Based on these facts, it is an intriguing project to investigate the ALP effect on the optical photon polarization state from blazars.
%This means that we have a better understanding of the initial state of photons. And blazars possess a relatively structured magnetic field compared to other astrophysical magnetic field, which can amplify the effects of ALPs. These facts motivate us to investigate the ALP effect on low energy photons from blazars in the optical band in the weak-mixing condition.

It is expected that ALPs would not significantly change the flux and linear polarization (LP) of the optical photons from blazars. However, we find that under some suitable conditions, ultra-light ALPs can lead to an appreciable circular polarization (CP), which may be an origin of optical CP in blazars. Conversely, the ALPs can be explored through some tentative CP observations of blazars.

In recent years, several optical polarization monitoring programs have been carried out, and mainly focus on the linear polarization signature \cite{Zhang:2019qgi}. However, the optical circular polarization in blazars has rarely been searched for. This is partly because that optical CP is expected to be low in usual physical processes \cite{rieger:2005possible}, requiring very sensitive instruments to deliver CP measurements. If the ALP-photon mixing can lead to an appreciable CP for the optical photons from blazars, we appeal to implement more sensitive experiments that can simultaneously measure the optical LP and CP to further nail down the effects of ALPs.

The paper is organized as follows. In Sec.~\ref{sec:mixing}, we present a brief calculation routine of the ALP-photon mixing and derive the CP formulas under the weak-mixing condition. In Sec.~\ref{sec:application}, we introduce the jet configuration of BL Lac, which is a type of blazar, and obtain results using the CP formulas. In Sec.~\ref{sec:possibleCP}, we analyze the ALP effects in the tentative CP observations of two blazars. In Sec.~\ref{sec:discussion}, we discuss several relevant factors that may affect the results. Conclusions are given in Sec.~\ref{sec:conclusions}. Note that all equations are expressed in Lorentz–Heaviside units throughout the paper.

\section{\label{sec:mixing}ALP-photon mixing in magnetic fields}
\subsection{Mixing in one domain}
The calculation routines used for the ALP-photon conversion have been described in detail in many papers, e.g., \cite{Mirizzi:2009aj,Bassan:2010ya,DeAngelis:2011id,Meyer:2014epa}. Nevertheless, for the reader's convenience, we briefly summarize all necessary elements here. The ALP-photon system can be described by the Lagrangian
\begin{equation}
  \begin{aligned}
    \mathcal{L}_\mathrm{ALP}&=\frac{1}{2} \partial^{\mu} a \partial_{\mu} a - \frac{1}{2} m_{a}^2 a^2 -\frac{1}{4} \gag
    aF_{\mu\nu}\tilde{F}^{\mu\nu}\\
    &=\frac{1}{2} \partial^{\mu} a \partial_{\mu} a - \frac{1}{2} m_{a}^2 a^2 + \gag a\bm{E}\cdot\bm{B},
  \end{aligned}
  \label{eq:lagrangian}
\end{equation}
where $a$ denotes the ALP field, $F_{\mu\nu}$ and $\tilde{F}^{\mu\nu}$ are the electromagnetic tensor and its dual, respectively. For the ALP, the mass $m_{a}$ is unrelated to its coupling with photons $\gag$. The ALP-photon mixing in the presence of an external magnetic field $\bm{B}$ is characterized by the $a\gamma\gamma$ vertex in the Lagrangian. Only the transverse component of the external magnetic field $\bm{B}_\mathrm{T}$ with respect to the direction of beam propagation matters.

We assume a monochromatic ALP-photon beam with energy of $E \gg m_a$ propagating in the $z$ direction. In the short-wavelength approximation, the equation of motion (EOM) can be written as \cite{Raffelt:1987im}
\begin{equation}
  \left(i \frac{d}{d z} + E + \mathcal{M} \right) \psi(z) = 0
  \label{eq:psiEOM}
\end{equation}
with
\begin{equation}
  \psi(z) = \left(\begin{array}{c}A_x (z) \\ A_y (z) \\ a (z) \end{array}\right),
\end{equation}
where $A_x(z)$ and $A_y(z)$ denote the LP amplitudes of the photon along the x- and y-axis, respectively, and $a(z)$ is the ALP amplitude.

In analysis, the astrophysical magnetic field is often supposed to be homogeneous in a single small domain. If its transverse component $\bm{B}_\mathrm{T}$ is set along with the $y$-axis, the mixing matrix is given by \cite{Kuster:2008zz}
\begin{equation}
  \mathcal{M}^{(0)} =  \left(
  \begin{array}{ccc}
    \Delta_{ \perp}  & 0 & 0 \\
    0 &  \Delta_{ \parallel}  & \Delta_{a \gamma}  \\
    0 & \Delta_{a \gamma} & \Delta_a
  \end{array}\right)~,
  \label{eq:MixingMatrix}
\end{equation}
with
\begin{gather}
  \Delta_{a\gamma} \equiv \frac{1}{2} \gag B_\mathrm{T},\label{eq:MixingTerm}\\
  \Delta_a \equiv - \frac{m_a^2}{2E}.\label{eq:MassTerm}
\end{gather}
$\Delta_{a\gamma}$ and $\Delta_a$ account for the ALP-photon mixing and ALP mass effect, respectively. Since we consider the photons in the optical band here, other off-diagonal $\Delta$-terms related to the Faraday rotation are neglected.

The diagonal terms related to the photon constitute four parts as \cite{Galanti:2018nvl,Davies:2020uxn}
\begin{gather}
  \Delta_\perp \equiv \Delta_\mathrm{pl} + \Delta_\mathrm{abs} + \Delta_\mathrm{CMB} + 2 \Delta_\mathrm{QED},\\
  \Delta_\parallel \equiv \Delta_\mathrm{pl} + \Delta_\mathrm{abs} + \Delta_\mathrm{CMB}+ 3.5  \Delta_\mathrm{QED},
\end{gather}
with
\begin{gather}
  \Delta_\mathrm{pl} \equiv -\frac{\omega^2_\mathrm{pl}}{2E},\label{eq:PlasmaTerm}\\
  \Delta_\mathrm{abs} \equiv \frac{i}{2\lambda_{\gamma}},\label{eq:AbsorptionTerm}\\
  \Delta_\mathrm{CMB} \equiv \chi_\mathrm{CMB}E,\label{eq:CMBTerm}\\
  \Delta_\mathrm{QED} \equiv \frac{\alpha E}{45 \pi} \left(\frac{B_\mathrm{T}}{B_\mathrm{cr}} \right)^2.\label{eq:QEDTerm}
\end{gather}
$\Delta_\mathrm{pl}$ accounts for the effective photon mass when the beam propagates in the plasma with the frequency parameter $\omega_\mathrm{pl}=(4\pi\alpha n_{e}/m_{e})^{1/2}$, where $\alpha$ is the fine-structure constant, $n_e$ and $m_e$ denote the electron number density and the electron mass, respectively. The other three contributions, including $\Delta_\mathrm{abs}$ the photon absorption term with the mean free path $\lambda_{\gamma}$, $\Delta_\mathrm{CMB}$ the photon dispersion term induced by the cosmic microwave background with the refractive index $\chi_\mathrm{CMB}\simeq0.522\times10^{-42}$ \cite{Dobrynina:2014qba}, and $\Delta_\mathrm{QED}$ the QED vacuum polarization term with
$B_\mathrm{cr}\simeq4.4\times10^{13}~\mathrm{G}$, are negligible in the scenario here. That is to say, we can safely make the approximation $\Delta_\parallel \approx \Delta_\perp \simeq \Delta_\mathrm{pl}$.

In a general situation, the transverse magnetic field $\bm{B}_{T}$ is not aligned with the y-axis. We denote $\phi$ the angle between $\bm{B}_{T}$ and the y-axis in a generic domain. Correspondingly, the mixing matrix can be obtained using the similarity transformation
\begin{equation}
  \mathcal{M} = V^{\dagger} (\phi) {\cal M}^{(0)} V (\phi)
\end{equation}
operated by the rotation matrix in the $x$-$y$ plane, namely
\begin{equation}
  V (\phi) =
    \left(
    \begin{array}{ccc}
      \cos\phi&-\sin\phi&0\\
      \sin\phi&\cos\phi&0\\
      0&0&1
    \end{array}
    \right).
\end{equation}

In view of our subsequent discussion, the polarization density matrix is introduced as
\begin{equation}
  \rho (z) =
  \left(\begin{array}{c}
      A_x (z) \\ A_y (z) \\ a (z)
  \end{array}\right)
  \otimes
  \left(\begin{array}{c}
      A_x (z)\  A_y (z)\ a (z)
  \end{array}\right)^{*},
\end{equation}
which obeys the Liouville-Von Neumann equation
\begin{equation}
  i \frac{d \rho}{d z} = [\rho, \mathcal{M}].
  \label{eq:rhoEOM}
\end{equation}
The solution of Eq.~(\ref{eq:rhoEOM}) is
\begin{equation}
  \rho (z) = \mathcal{T}(z,z_0) \, \rho (z_0) \, \mathcal{T}^{\dagger}(z,z_0)~,
  \label{eq:rhosol}
\end{equation}
where the transfer function $\mathcal{T}(z,z_0)$ is the solution of the EOM in the form of $\psi(z)=\mathcal{T}(z,z_0)\psi(z_0)$ with the initial condition $\mathcal{T}(z_0,z_0) = 1$.
The probability that an ALP-photon beam initially in the state $\rho(z_0)$ converts into the state $\rho_\mathrm{f}$ at a position with $z$ can be computed as
\begin{equation}
  P=\mathrm{Tr}(\rho_\mathrm{f}\mathcal{T}(z,z_0) \, \rho (z_0) \, \mathcal{T}^{\dagger}(z,z_0)).
\end{equation}

For a fixed magnetic field, the probability that a photon polarized along the y-axis oscillates into an ALP after a distance $d$ can simply read
\begin{equation}
  P_{\gamma \to a}=\cos^{2}\phi\sin^{2}(2\theta)\sin^{2}(\frac{\Delta_\mathrm{osc}d}{2}),
  \label{eq:probability}
\end{equation}
with the ALP-photon mixing angle
\begin{equation}
  \theta=\frac{1}{2}\arctan(\frac{2\Delta_{a\gamma}}{\Delta_\parallel-\Delta_a})
\end{equation}
and the oscillation wave number
\begin{equation}
  \Delta_\mathrm{osc}=[(\Delta_a-\Delta_\parallel)^{2}+4\Delta_{a\gamma}^{2}]^{1/2}.
\end{equation}

As for the numerical calculation with a more realistic configuration of the astrophysical magnetic field, the field environment can be sliced into $N$ consecutive domains with homogeneous magnetic fields. For each domain, the EOM can be solved exactly. By iterating, we can derive the photon and/or axion state at any propagation distance.

\subsection{Optical circular polarization}

The $2 \times 2$ density matrix of the photon polarization (i.e., the 2-2 block of the density matrix for the ALP-photon system) can be expressed in terms of the Stokes parameters \cite{Kosowsky:1994cy}
\begin{equation}
  \rho_{\gamma} = \frac{1}{2} \left(
    \begin{array}{cc}
      I + Q & U - i V \\
      U + i V & I - Q\\
    \end{array}
  \right).
\end{equation}
The degree of CP is defined as \cite{Rybicki:2004hfl}
\begin{equation}
  \Pi_{C}  \equiv \frac{V}{I} .
\end{equation}

In this study, we work within the weak-mixing condition, i.e.
\begin{equation}
  \begin{gathered}
    |\Delta_\mathrm{pl}| \gg \Delta_{a\gamma},\\
    |\Delta_\mathrm{pl}| \gg |\Delta_a|.
  \end{gathered}
  \label{eq:condition}
\end{equation}
In this regime, $\theta\ll1$ and the conversion probability turns out to be vanishingly small. Expanding Eq.~(\ref{eq:rhosol}) and neglecting the high order terms, the evolution of the Stokes parameters in a single domain region can be obtained as
\begin{gather}
  I(z) = I(z_{0}) - \mathcal{I} (1-\cos\kappa),\\
  Q(z) = Q(z_{0}) - \mathcal{Q}_{1}(1-\cos\kappa) - \mathcal{Q}_{2} \sin\kappa,\\
  U(z) = U(z_{0}) - \mathcal{U}_{1}(1-\cos\kappa) - \mathcal{U}_{2} \sin\kappa,\\
  V(z) = V(z_{0}) \cos \kappa + \mathcal{V} \sin \kappa,
\end{gather}
where
\begin{gather}
  \mathcal{I} \equiv Q(z_{0})\sin^{2}2\phi,\\
  \mathcal{Q}_{1} \equiv \frac{1}{2}U(z_{0})\sin4\phi,\\
  \mathcal{Q}_{2} \equiv V(z_{0})\sin2\phi,\\
  \mathcal{U}_{1} \equiv \left[Q(z_{0})\sin2\phi + U(z_{0})\cos2\phi\right]\cos2\phi,\\
  \mathcal{U}_{2} \equiv V(z_{0})\cos2\phi,\\
  \mathcal{V} \equiv Q(z_{0})\sin 2\phi + U(z_{0}) \cos 2\phi,
  \label{eq:MathcalV}\\
  \kappa \equiv \frac{\Delta_{a\gamma}^{2}}{\Delta_\mathrm{pl}-\Delta_a}(z-z_{0})~.
\end{gather}

Supposing $|\kappa|\ll1$ in each domain and the degree of CP of the incident beam can be neglected, the above equations show that the changes of $I$, $Q$, and $U$ are negligible at first leading order. For the CP, we can evaluate it across multidomains as
\begin{equation}
  \begin{aligned}
    &V(z) - V(z_{0})\\
    \approx & \int_{z_{0}}^{z} \mathcal{V} \frac{\Delta_{a\gamma}^{2}}{\Delta_\mathrm{pl}-\Delta_a} dz^{\prime}\\
    = & -\frac{m_{e} \gag^{2} \mathcal{V} E}{8\pi\alpha}  \int_{z_{0}}^{z}  \frac{B_{T}^{2}(z^{\prime})}{n_{e}(z^{\prime})} dz^{\prime}.
  \end{aligned}
  \label{eq:Vsol}
\end{equation}

The parameter $\mathcal{V}$ is determined by the LP state of incident photons and the direction of the transverse magnetic field. For the above equation, we have made the assumption that the direction of transverse magnetic field remains unchanged, thus $\mathcal{V}$ can be treated as constant. A general polarization density matrix of a beam of linearly polarized photons can be written as
\begin{equation}
  \begin{gathered}
      \rho_\gamma(\Pi_\mathrm{L},\psi)= \left(
      \begin{array}{cc}
        \frac{1-\Pi_\mathrm{L}\cos2\psi}{2} & \frac{\Pi_\mathrm{L}\sin2\psi}{2} \\
        \frac{\Pi_\mathrm{L}\sin2\psi}{2} & \frac{1+\Pi_\mathrm{L}\cos2\psi}{2}\\
      \end{array}
    \right),
  \end{gathered}
  \label{eq:densityeq}
\end{equation}
where
\begin{equation}
    \Pi_\mathrm{L}\equiv\frac{\sqrt{Q^2+U^2}}{I}
\end{equation}
is the degree of LP, and $\psi$ is the polarization angle relative to the y-axis. Then, Eq.~(\ref{eq:MathcalV}) can be rewritten as
\begin{equation}
  \mathcal{V}=\Pi_\mathrm{L}\sin2(\phi-\psi).
  \label{eq:MathcalV1}
\end{equation}
The above equations show that the magnitude of the ALP induced CP is proportional to the LP degree of the photons in the weak-mixing regime. When $\phi-\psi$, which denotes the angle between the LP and the transverse magnetic field, equals $(2k+1)\pi/4$ with $k\in\mathbb{Z}$, $\mathcal{V}$ reaches the extreme value $\pm\Pi_\mathrm{L}$.

Even though the energy of the photon and the magnetic field strength may not be high, the integration indicates that the ALPs can induce a considerable CP as long as the field environment is suitable and the spatial scale of magnetic field is large enough. This makes blazars good candidate sources, as discussed in the next section.

\section{\label{sec:application}Application in blazars}
\textit{Blazars} are active galactic nuclei with a relativistic jet pointing very close to Earth \cite{Urry:1995mg}. The highly collimated and Doppler boosted jet make them bright and variable in all wavebands from the radio to $\gamma$-ray \cite{Angel:1980yg}. Blazars are categorized as BL Lac objects (BL Lacs) \cite{Shaw:2013pp} and flat-spectrum radio quasars (FSRQs) \cite{Shaw:2012aq}, which were mostly called optically violent variable quasars. Compared to BL Lacs, FSRQs are generally more luminous and show broad emission lines in the optical and UV bands. The jet of BL Lacs could extend to $\sim\mathrm{kpc}$. In the jet the magnetic field is substantially ordered and predominantly traverse to the jet \cite{Pudritz:2012xj}.

The degree of optical LP in blazars is typically at a level of $\sim10\%$ and can be up to $\sim50\%$ sometimes, while no definite CP has been detected\cite{Hutsemekers:2010fw,Zhang:2019qgi}. Therefore, we can set a constraint on the parameters of the ALP utilizing the effect discussed in the previous section. We focus on BL Lacs in the discussion below, considering that the field environment of FSRQs are more complicated and the geometry and intensity of $\bm{B}$ are less clear compared to BL Lacs \cite{Pudritz:2012xj,Prandini:2022khy}.

\subsection{\label{subsec:BLLacs}Field configuration of BL Lacs}
Faraday rotation measure (RM) is an important tool in astronomy, and enables one to study the magnetic field and electron number density of blazar jets. Many observations have detected the transverse RM gradients across the jets of BL Lacs, which evidence that these jets have the toroidal or helical magnetic fields \cite{Gabuzda:2004kc,gabuzda:2017parsec,Kravchenko:2017aun}. The conservation of the Poynting flux demands that the magnetic field strength varies as $d^{-1}$, where $d$ is the transverse size of the jet \cite{Begelman:1984mw,rees:1987magnetic}. The conservation of particles demands that the number density varies as $r^{-2}$ \cite{OSullivan:2009dsx}, where $r$ is the distance from the source of the jet. In this work, we adopt the conical shaped model used in \cite{Meyer:2014epa,Tavecchio:2014yoa}. In the comoving frame, the transverse magnetic field and electron density of the jet are modeled by
\begin{equation}
  B_T^\mathrm{jet} = B_{T0} \left(\frac{r}{r_\mathrm{E}}\right)^{-1}
\end{equation}
and
\begin{equation}
  n_e^\mathrm{jet} = n_{e0} \left(\frac{r}{r_\mathrm{E}}\right)^{-2}
\end{equation}
for $r>r_\mathrm{E}$, where $r_\mathrm{E}$ is the distance of the emission site from the central black hole. The value of $r_\mathrm{E}$ is usually inferred from the size of the emission region and the jet aperture angle, typically ranging from $0.01~\mathrm{pc}$ to $0.1~\mathrm{pc}$ \cite{Celotti:2007rb}. However, it will be found in the latter discussion that the value of $r_\mathrm{E}$ has little effect on our results, thus we choose $r_\mathrm{E}=0.05~\mathrm{pc}$ in order to be definite. For the electron density at the emission site $n_\mathrm{e0}$ and the size of the jet $r_\mathrm{max}$, the typical values are $n_\mathrm{e0}\simeq5\times10^{4}~\mathrm{cm}^{-3}$ and $r_\mathrm{max}\simeq1~\mathrm{kpc}$ \cite{Meyer:2014epa,Tavecchio:2014yoa}. The exact value of $n_\mathrm{e0}$ for the specific blazar is not definite in general, however it can be inferred from fitting parameters of the spectral energy distributions (SEDs) of blazars.

The SEDs of blazars are dominated by two distinct radiative components: a broad low-frequency component from radio to optical/UV or x-ray and a high-frequency component from x-ray to $\gamma$-ray. It is generally accepted that the low-frequency component is produced by the synchrotron radiation of relativistic electrons in the jets. For the interpretation of the high-frequency component,
synchrotron self-Compton (SSC) and external Compton (EC) are two main mechanisms. Hence, the popular models differ by the radiation particles into two kinds, namely leptonic (emission dominated by electrons and possibly positrons) and hadronic (emission dominated by protons) models \cite{blandford:2019relativistic,Prandini:2022khy}. In the leptonic models, the SED fitting usually gives a magnetic field strength of $0.1-1~\mathrm{G}$ for Intermediate BL Lacs and $1-3~\mathrm{G}$ for Low-Frequency-Peaked BL Lacs \cite{Boettcher:2013wxa}. However, in the hadronic models, the magnetic field strength is much lager, ranging from $10~\mathrm{G}$ to $100~\mathrm{G}$ \cite{Boettcher:2013wxa}. For the convenience of study, we choose the typical parameter $B_{T0}\simeq1~\mathrm{G}$. The different results can be inferred depending on the radiation models of blazars.

The current most robust bound on $\gag$ in the parameter region of interest here is provided by the CAST experiment, which constrains $\gag<6.6\times10^{-11}~\mathrm{GeV}^{-1}$ for $m_a<0.02~\mathrm{eV}$ \cite{CAST:2017uph}. Other studies can also set stronger constraints depending on specific procedures (e.g., \cite{Grifols:1996id,Payez:2014xsa,Berg:2016ese,Marsh:2017yvc}). In this work, we choose $\gag=5\times10^{-11}~\mathrm{GeV}^{-1}$ allowed by the CAST bound as the benchmark value.

In the rest of this subsection, we discuss the validity of the weak-mixing condition. The first condition of Eq.~(\ref{eq:condition}) requires
\begin{widetext}
\begin{equation}
  \left(\frac{\gag}{5\times10^{-11}\mathrm{GeV}^{-1}}\right)\left(\frac{E}{\mathrm{1eV}}\right)\ll1.45\left(\frac{B_{T}}{1\mathrm{G}}\right)^{-1}\left(\frac{n_e}{1\mathrm{cm}^{-3}}\right).
\end{equation}
\end{widetext}

In the case of blazar, this means that the weak-mixing condition is satisfied for
\begin{widetext}
\begin{equation}
    r\ll28~\mathrm{kpc}\left(\frac{B_{T0}}{1~\mathrm{G}}\right)^{-1}\left(\frac{n_{e0}}{5\times10^{4}~\mathrm{cm}^{-3}}\right)\left(\frac{r_\mathrm{E}}{0.05~\mathrm{pc}}\right)\left(\frac{\gag}{5\times10^{-11}~\mathrm{GeV}^{-1}}\right)^{-1}\left(\frac{E^\prime}{2~\mathrm{eV}}\right)^{-1}\left(\frac{\delta}{15.64}\right).
\end{equation}
\end{widetext}
where $\delta$ is the relativistic Doppler factor and $E^\prime$ is the energy of photons in the rest frame. The photon energy $E$ in the co-moving frame is related to the observed energy $E^\prime$ through the Doppler factor $\delta$ as $E=E^\prime/\delta$. The observed energy $E^\prime\simeq2~\mathrm{eV}$ corresponds to the photons in the optical waveband, and $\delta=15.64$ is taken to be the same typical value as \cite{Meyer:2014epa}. Straightforwardly, the more stringent constraint on $\gag$ is required for the larger magnetic field strength to meet the condition above.

The second condition of Eq.~(\ref{eq:condition}) requires the mass of the ALP and the size of the jet
\begin{equation}
  m_{a}\ll 3.8\times10^{-11}\mathrm{eV}\left(\frac{n_{e}}{1\mathrm{cm}^{-3}}\right)^{1/2}.
  \label{eq:MassCondition}
\end{equation}
Thus, the ALPs considered here need to be very light as $m_{a}\lesssim 10^{-13}~\mathrm{eV}$, in order to satisfy the weak-mixing condition through the jet region, i.e., $r \lesssim r_\mathrm{max}$.

\subsection{Results}\label{subsec:result}
Given the magnetic field configuration of BL Lacs and the parameters of ALPs, we obtain
\begin{widetext}
\begin{equation}
  \begin{aligned}
    &V(r) - V(r_\mathrm{E}) \\
    =& -\frac{m_{e} \gag^{2} \mathcal{V}}{8\pi\alpha}  \frac{E^\prime}{\delta} \frac{B_{T0}^{2}}{n_{e0}}(r-r_\mathrm{E})\\
    \simeq &-0.134\mathcal{V}\left(\frac{\gag}{5\times10^{-11}~\mathrm{GeV}^{-1}}\right)^{2}\left(\frac{E^\prime}{2~\mathrm{eV}}\right)\left(\frac{\delta}{15.64}\right)^{-1}\left(\frac{B_{T0}}{1~\mathrm{G}}\right)^{2}\left(\frac{n_\mathrm{e0}}{5\times10^{4}~\mathrm{cm}^{-3}}\right)^{-1}\left(\frac{r}{1~\mathrm{kpc}}\right).
  \end{aligned}
  \label{eq:blazarVsol}
\end{equation}
\end{widetext}

In some cases, we should modify the above equation, since the value of $n_\mathrm{e0}$ is not explicit in blazar observation, as we mentioned before. In the SED model, a power-law distribution of relativistic electrons and/or pairs with low and upper energy cutoffs at $\gamma_{1}$ and $\gamma_{2}$, respectively, and the power-law index $q$ is presumed generally \cite{Boettcher:2013wxa,boettcher:2002x}. In order to infer the electron density from the SED fitting, we assume a broken power-law distribution of electrons with index $p$ for nonrelativistic electrons \cite{Maraschi:2002pp},
\begin{equation}
    n_\mathrm{e}(\gamma)=\left\{\begin{array}{cc}
        n_\mathrm{e0}\gamma^{-p} & \mathrm{for}~1\leqslant\gamma<\gamma_{1} \\
        n_\mathrm{e0}\gamma_{1}^{q-p}\gamma^{-q} & \mathrm{for}~\gamma_{1}\leqslant\gamma\leqslant\gamma_{2}\\
        0&\mathrm{else}\\
      \end{array}
      \right.,
\end{equation}
where $\gamma$ is the Lorentz factor of the electrons. The index $p$ is chosen to be the typical value of $2$ hereafter \cite{Tavecchio:2009zb}. In addition, assuming $B_\mathrm{T0}\approx B_0$ at the emission region, we can modify Eq.~(\ref{eq:blazarVsol}) with
\begin{equation}
    \left(\frac{B_{T0}}{1~\mathrm{G}}\right)^{2}\left(\frac{n_\mathrm{e0}}{5\times10^{4}~\mathrm{cm}^{-3}}\right)^{-1}\approx \epsilon_{Be} f_\gamma,
\end{equation}
where $\epsilon_{Be}\equiv L_{B}/L_{e}$ is the ratio of the power carried in the magnetic field and the kinetic power in relativistic electrons, and
\begin{equation}
    f_\gamma \equiv \frac{\gamma_{2}^{2-q}-\gamma_{1}^{2-q}}{(2-q)\gamma_{1}^{p-q}}
\end{equation}
accounts for the factor related to the electron density.

\begin{figure}[htbp]
  \includegraphics[width=1.0\columnwidth]{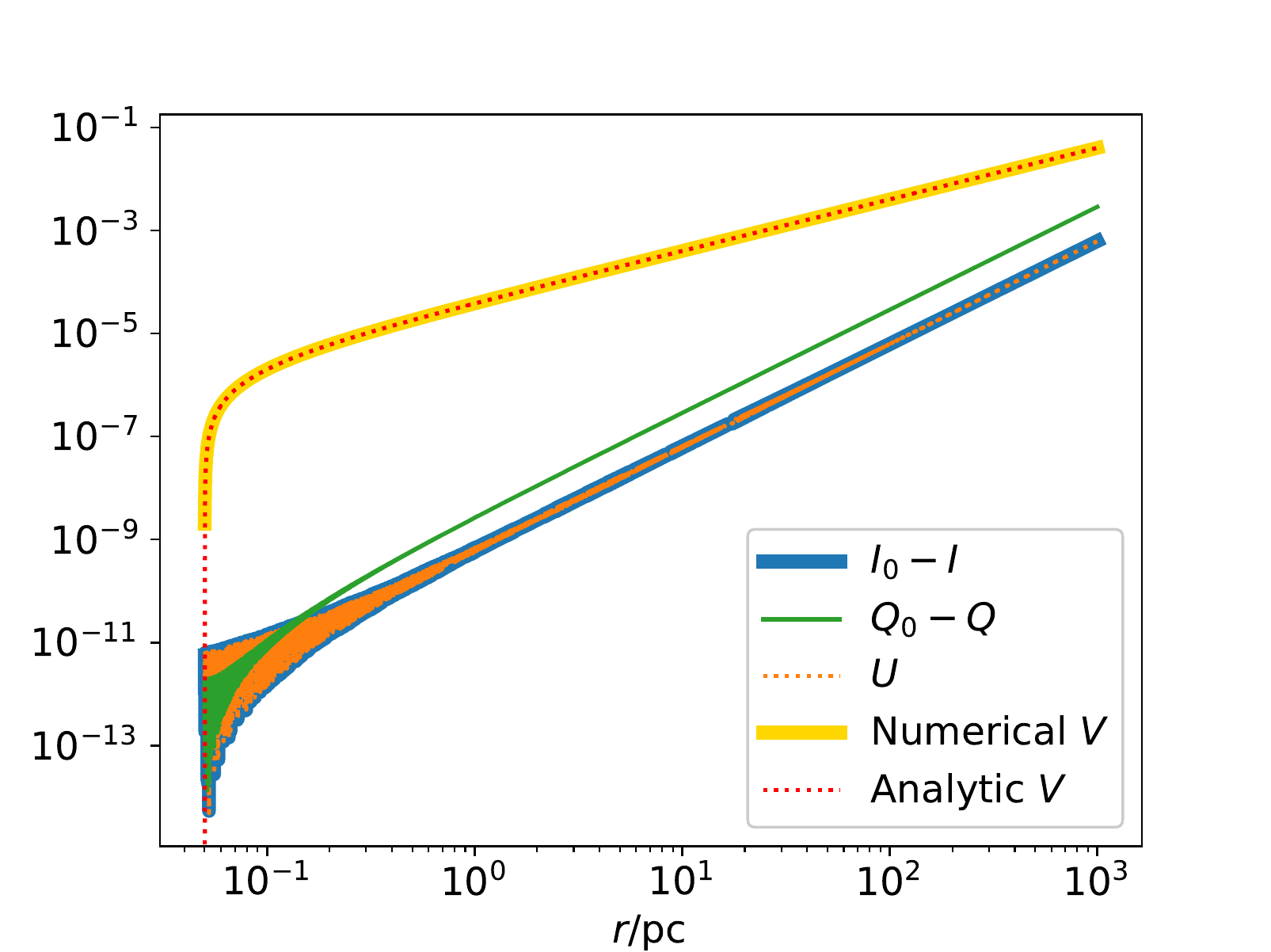}
  \caption{\label{fig:illustration} The evolution of Stokes parameters through the jet with the distance $r$. The $Q_0$ and $I_0$ represent the initial value at $r_\text{E}$. Relevant parameters are set to be their typical value while $\Pi_L=0.3$ and $m_{a}=10^{-15}~\mathrm{eV}$. The value of $\mathcal{V}$ is chosen to be equal to $-\Pi_L$ in order to maximize the ALP effect.}
\end{figure}

By solving Eq.~(\ref{eq:rhoEOM}) numerically, we can obtain the evolution of the Stokes parameters with distance, as shown in Fig.~\ref{fig:illustration}. Without loss of generality, we suppose the initial photons linearly polarize along the y-axis, i.e., $\psi=0$, with a polarization degree $\Pi_L=0.3$. In this case, the polarization density matrix is $\rho_{0}=\mathrm{diag}(0.65,0.35,0)$ (see Eq.\eqref{eq:densityeq}). We assume that the direction of the transverse magnetic field remains unchanged along the jet for simplicity. It is chosen to be $\phi=3\pi/4$, which maximizes the effect of ALPs and leads to $\mathcal{V}=-\Pi_L=-0.3$. We take $m_{a}=10^{-15}~\mathrm{eV}$ and set other relevant parameters as their typical values. The length of domains along $z$ are determined such that the magnetic field decreases by the decrease ratio $\mathcal{R}$ from one domain to the next, i.e., $\mathcal{R}=1-B_{i+1}/B_{i}$ with the subscription denoting the $i$th domain. In Fig.~\ref{fig:illustration}, we have set $\mathcal{R}=0.001$.
%larger ratio would obtain slightly lager (smaller) CP since the magnetic field in each domain would be overestimated (underestimated).

From Fig.~\ref{fig:illustration}, we can see that the analytic result of $V$ derived from Eq.~(\ref{eq:blazarVsol}) is consistent with the numerical result. This justifies our analytic formulas.
Another interesting feature is that the ALPs can induce a much larger CP compared with LP. Here we give a general qualitative analysis. As long as $P_{\gamma\to a}$ is small, the LP generated by dichroism can be approximated to be the conversion probability, i.e., Eq.~(\ref{eq:probability}) \cite{Jain:2002vx,Hutsemekers:2010fw}
\begin{equation}
  \Delta\Pi_\mathrm{L}\simeq P_{\gamma\to a} \approx \sin^{2}(2\theta)\sin^{2}(\xi/2),
  \label{eq:DeltaLP}
\end{equation}
where $\xi=L/l_\mathrm{osc}$ is the ratio of the coherent length of magnetic field to oscillation length. The photons acquire a polarization-dependent phase shift (retardance)
\begin{equation}
\phi_{a}\simeq\sin^{2}(2\theta)(\xi-\sin\xi)
\end{equation}
in the mixing with ALPs, which results in CP. If $\phi_{a}\ll1$, then the CP induced by phase shift can be expressed as
\begin{equation}
  V=Q_{0}\sin\phi_{a}\simeq Q_{0}\sin^{2}(2\theta)(\xi-\sin\xi)
  \label{eq:EstimateV}
\end{equation}
with adopting the convention of $U_{0}=0$ and $Q_{0}>0$. For the BL Lacs considered here, the corresponding oscillation length is
\begin{equation}
  l_\mathrm{osc}\approx|\Delta_\mathrm{pl}|^{-1}\approx9\times10^{-4}\mathrm{pc}\left(\frac{E}{\mathrm{1eV}}\right)\left(\frac{n_e}{1\mathrm{cm}^{-3}}\right)^{-1}.
\end{equation}
It is obvious that the oscillation length is far less than the scale of the jet $l_\mathrm{osc}\ll r_\mathrm{max}$, supposing the magnetic field is coherent over the entire jet region. Thus, the CP produced by the mixing is much larger than that of LP. Besides, Eq.~(\ref{eq:DeltaLP}) can also explain the oscillatory behavior of $I$ and $U$ in Fig.~\ref{fig:illustration}.

\begin{figure}[htbp]
  \includegraphics[width=1.0\columnwidth]{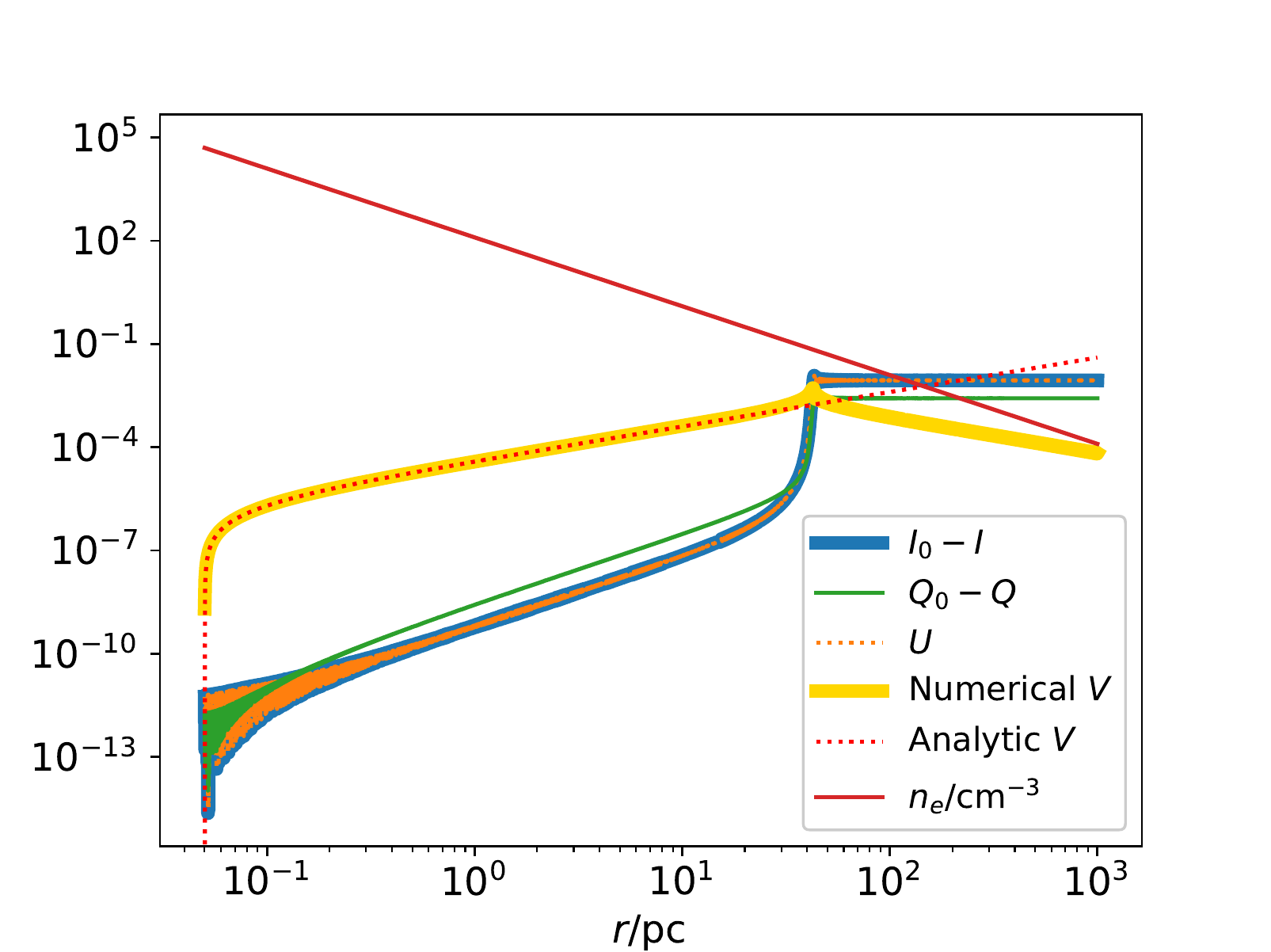}
  \caption{\label{fig:condition} Similar to Fig.~\ref{fig:illustration} except that $m_{a}=10^{-11}~\mathrm{eV}$. The red solid line represents the value of electron number density in unit of $\mathrm{cm}^{-3}$.}
\end{figure}

In Fig.~\ref{fig:condition} we demonstrate a similar case, in which all parameters remain unchanged except that $m_{a}=10^{-11}~\mathrm{eV}$. We find that when $n_\mathrm{e}$ approaches $1~\mathrm{cm}^{-3}$, crossing the turning point where the resonant conversion occurs, the behavior of mixing is totally different. This is because the weak-mixing condition Eq.~(\ref{eq:MassCondition}) is no longer satisfied, and hence Eq.~(\ref{eq:blazarVsol}) is not applicable.

A few attempts have been made to measure optical CP in blazars \cite{Hutsemekers:2010fw,valtaoja:1993optical,wagner:2001circular}, however no definite optical CP in blazars has been detected. Hutsemékers \textit{et al.} \cite{Hutsemekers:2010fw} reported null detection of CP with typical uncertainties $<0.1\%$ in 21 quasars, except for two highly polarized blazars. There has not been a confident detection of optical CP in BL Lacs so far. However, some observations still reported tentative results of optical CP in some sources, such as the blazar 3C 66A \cite{takalo:1993simultaneous,tommasi:2001multiband}. We will discuss the ALP interpretation of these observations in the next section.

\begin{figure}[htbp]
  \includegraphics[width=1.0\columnwidth]{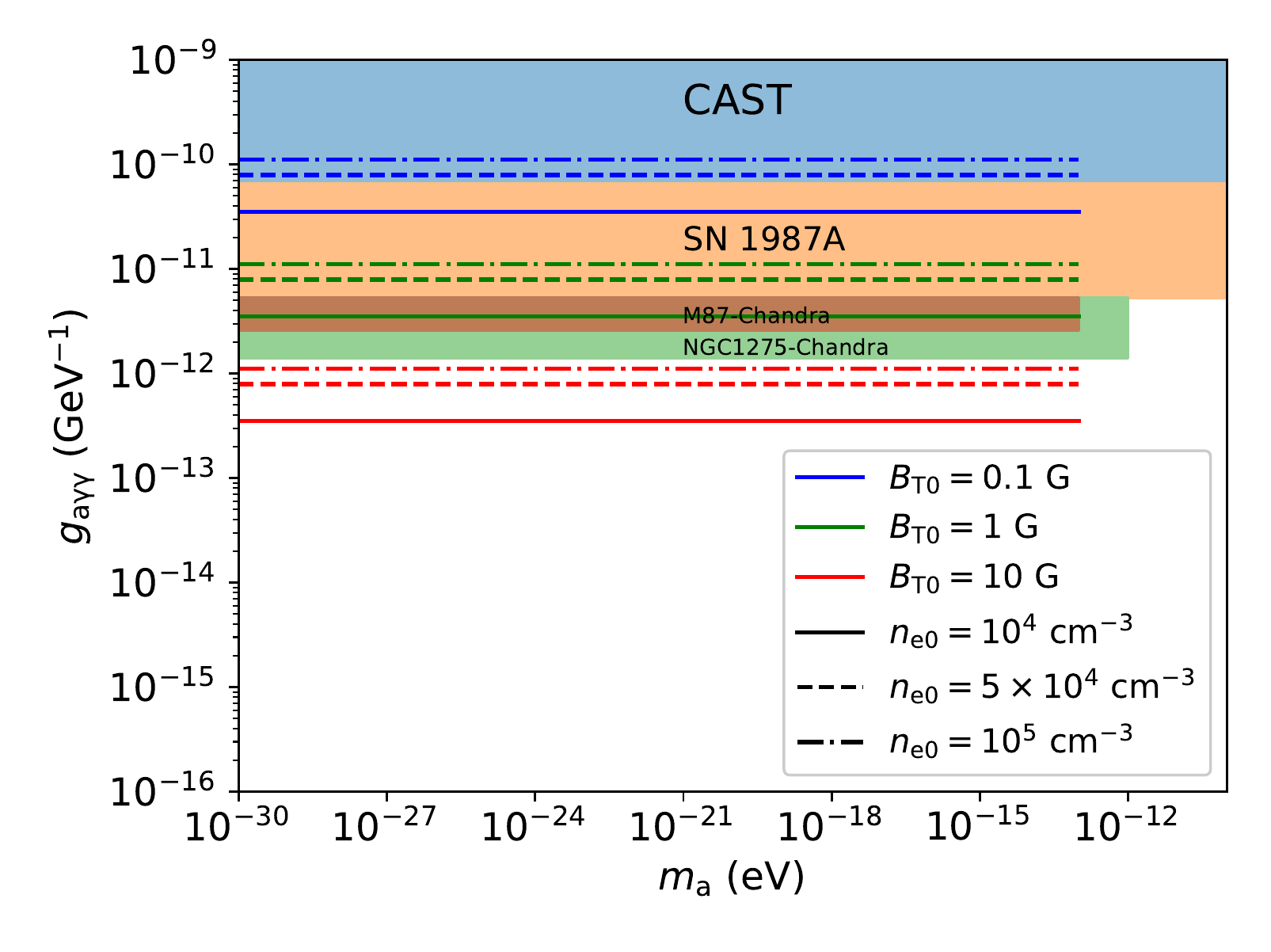}
  \caption{\label{fig:constraints} Bounds on the ALP-photon coupling in comparison to previously obtained bounds \cite{CAST:2017uph,Payez:2014xsa,Berg:2016ese,Marsh:2017yvc}.}
\end{figure}

Considering a detection sensitivity at the level of $\sim0.1\%$, Eq.~(\ref{eq:blazarVsol}) can directly give a constraint of $\gag\cdot B_\mathrm{T0}\lesssim7.9\times10^{-12}~\mathrm{G\cdot GeV}^{-1}$, i.e., $\gag\lesssim7.9\times10^{-12}~\mathrm{GeV}^{-1}$ for $B_\mathrm{T0}=1~\mathrm{G}$. Compared to leptonic models, hadronic models posses a higher magnetic field strength and less electron density, and thus can give more stringent constraints on the ALP-photon coupling. For example, $B_\mathrm{T0}\sim10~\mathrm{G}$ would require $\gag\lesssim7.9\times10^{-13}~\mathrm{GeV}^{-1}$ assuming the same electron density.
%In other words, if light ALPs are found, the radiation models with high magnetic field strength could be excluded.
In Fig.~\ref{fig:constraints}, we demonstrate the constraints on $g_a\gamma$ for different choices of $B_\mathrm{T0}$ and $n_\text{e0}$. For comparison, some previous bounds in the literature are also included \cite{CAST:2017uph,Payez:2014xsa,Berg:2016ese,Marsh:2017yvc}\footnote{There are also many studies on the varying polarization signal induced by the oscillating ALP field, which can set very stringent bounds in part of the parameter space of interest here (See, e.g., \cite{Chen:2021lvo,Castillo:2022zfl,Dessert:2022yqq,SPT-3G:2022ods}). As these results base on the assumption that the ALP dark matter exists near sources, which is not necessary in this analysis, they are not included in Fig.~\ref{fig:constraints}.}.

It is worth noting that the above constraints are derived for the optimal case $\mathcal{V}=\pm\Pi_\mathrm{L}$, while the value of $\mathcal{V}$ is supposed to be smaller in the more realistic case. More detailed discussions on this issue are given in Sec.~\ref{subsec:JetStructure}. The results obtained here are only rough estimation, which are limited by the knowledge of the jet properties as well as the measurement precision of optical CP. There is no doubt that further research of BL Lacs and improvement of future CP measurements in the optical band will significantly improve our results.

\section{\label{sec:possibleCP}ALP interpretation of the tentative CP observation}
Although no definite optical CP in blazars has been detected, some observations reported tentative optical CP results in a few particular objects. For example, in 2001, Tommasi \textit{et al.} \cite{tommasi:2001multiband} reported a marginal detection of optical CP at $2\sigma$ level in V and R bands for 3C 66A by using the Nordic Optical Telescope, which gives an upper limit of $\Pi_\mathrm{C} < 0.4\%$ at $3\sigma$ confidence level. Besides, a $3-6\sigma$ detection of CP with larger values for 3C 66A was also claimed by Takalo and Sillanpaa \cite{takalo:1993simultaneous}, using the same telescope in 1993. Moreover, Hutsemékers \textit{et al.} \cite{Hutsemekers:2010fw} reported small but significant optical CP in two blazars with uncertainties $<0.1\%$. These tentative observations could be interpreted by the ALPs in the context. Noting that the two blazars in \cite{Hutsemekers:2010fw} are not BL Lacs, the model adopted in Sec.~\ref{subsec:BLLacs} may not be applicable. Therefore, we put these two objects aside and focus on the 3C 66A at first. Besides, considering that the state of LP was uncertain and the amount of data is inadequate in the research of Tommasi \textit{et al.} \cite{tommasi:2001multiband}, we only analyze the data given by Takalo and Sillanpaa \cite{takalo:1993simultaneous}.

\begin{figure}[htbp]
  \includegraphics[width=1.0\columnwidth]{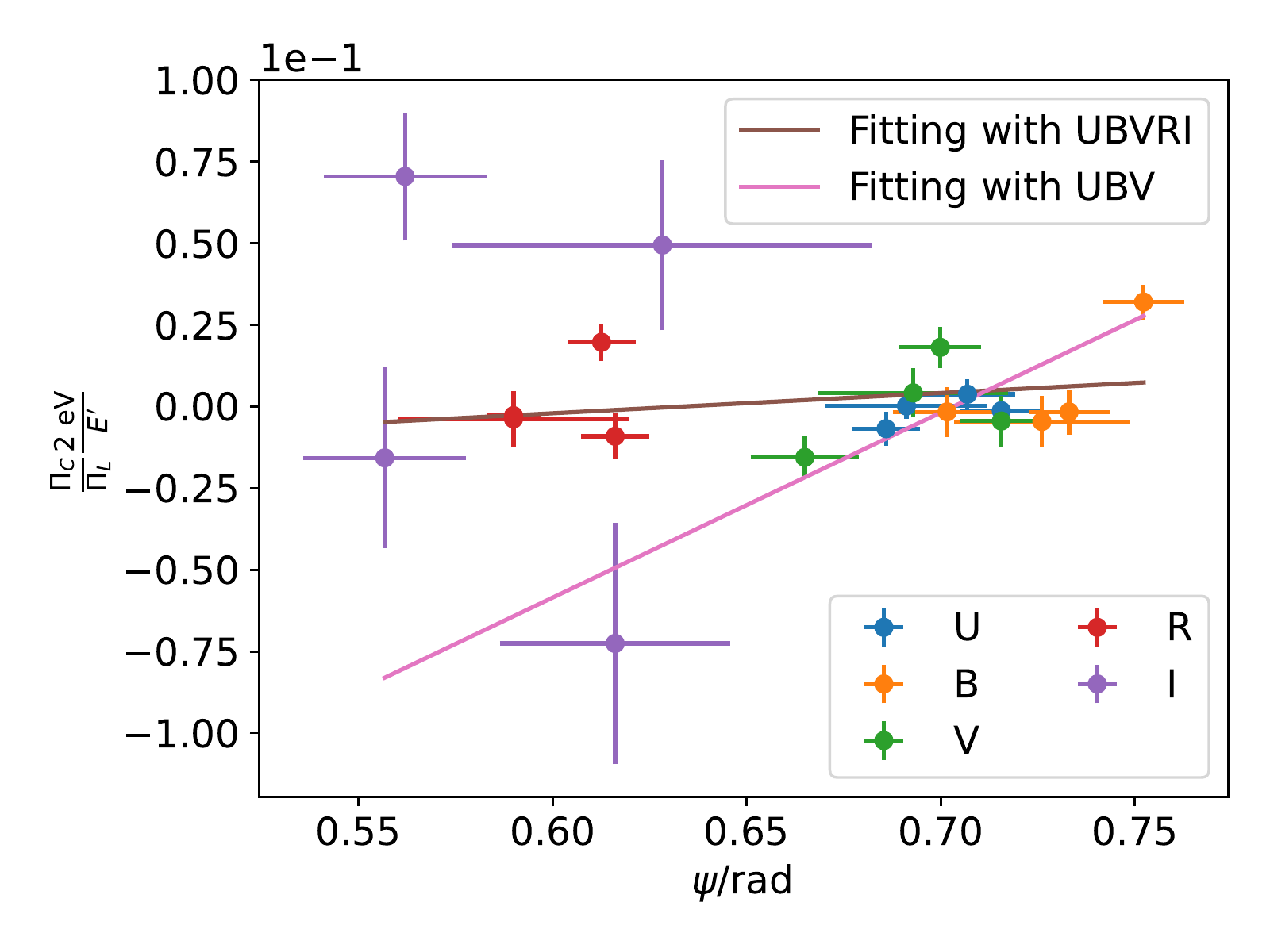}
  \caption{\label{fig:3C66A1993}Linear polynomial fit between the polarization angle $\psi$ and $\frac{\Pi_\mathrm{C}}{\Pi_\mathrm{L}}\frac{2~\mathrm{eV}}{E^\prime}$, using polarization observation data of 3C 66A in \cite{takalo:1993simultaneous}. The letters UBVRI represent different band filters.}
\end{figure}

The handness of CP changed during the measurement in \cite{takalo:1993simultaneous}. This implies that $\phi-\psi\approx k\pi/2$ with $k\in\mathbb{Z}$ in the ALP scenario. Hence, we can make the approximation $\mathcal{V}\approx2\Pi_\mathrm{L}(\phi-\psi)$ in Eq.~(\ref{eq:Vsol}) and obtain
\begin{equation}
    \frac{\Pi_\mathrm{C}}{\Pi_\mathrm{L}E^\prime}\propto\gag^{2}\epsilon_{Be}f_{\gamma}\delta^{-1}r_\mathrm{max}(\psi-\phi).
\end{equation}
The observations were made in the UBVRI photometric systems. Taking into account the effective energy of the color bands, we implement a linear polynomial fit between $\Pi_\mathrm{C}/(\Pi_\mathrm{L}E^\prime)$ and the polarization angle $\psi$. The results are shown in Fig.~\ref{fig:3C66A1993} and Table.~\ref{tab:3C66Afit}. Here, we have supposed that the orientation of the magnetic field is fixed. Owing to the small value of the CP, the quality of our analysis is sensitive to the accuracy of the polarization angle. It can be seen from Fig.~\ref{fig:3C66A1993} that the data points of RI bands are distributed differently from those of UBV bands. This may be attributed to the systematic differences caused by the two kinds of photomultiplier tubes in the UBV and RI channels. Physical processes that result in the energy dependent rotation of the polarization angle can also affect the distribution. We consider two cases here: one includes the RI bands, while the other not. The potential rotation of the photon polarization angle generated by other effects in the propagation through the astronomical space is neglected in the fitting.

\begin{table}[htbp]
\begin{tabular}{ccccc}
\hline
\hline
\addlinespace[0.3ex]
~~Bands~ & ~~~$a_0$~~ & ~~$\sigma_{a_0}$~~ & ~~$a_1$~~ & ~~$\sigma_{a_1}$~~\\
\addlinespace[0.3ex]
\hline
\addlinespace[0.5ex]
UBVRI & $-0.39$ & 0.45 & 0.62 & 0.66 \\
\addlinespace[0.3ex]
UBV & $-4.0$ & 1.1 & 5.7 & 1.6 \\ \hline
\end{tabular}
\caption{\label{tab:3C66Afit}Coefficients of linear polynomial fit to observation in 3C 66A in form of $\frac{\Pi_\mathrm{C}}{\Pi_\mathrm{L}}\frac{2~\mathrm{eV}}{E^\prime}=a_0+a_1\psi$. Parameters are in unit of $10^{-1}$.}
\end{table}

As shown in Table.~\ref{tab:3C66Afit}, no clear physical conclusion can be drawn from the fit in the UBVRI case, due to the large standard error compared to the corresponding fitting parameters. In the UBV case, the direction of the transverse magnetic field of the jet $\phi$ is estimated to be $0.70~\mathrm{rad}$ (40.3\textdegree) or $3.84~\mathrm{rad}$ (220.4\textdegree). From the slope of the line, we obtain
\begin{equation}
\gag\approx7\times10^{-11}~\mathrm{GeV}^{-1}\left[\epsilon_{Be}f_\gamma\left(\frac{15.64}{\delta}\right)\left(\frac{r_\mathrm{max}}{1~\mathrm{kpc}}\right)\right]^{-1/2}.
\end{equation}

The jet model fitting to the observed SED and optical variability patterns of 3C 66A usually gives $\delta\sim30$ and $f_\gamma\approx1$ \cite{Abdo:2010dsc,Boettcher:2013wxa,Reimer:2009zzc}. In the leptonic models, a pure SSC model would require a tiny equipartition ratio $\epsilon_{Be}$ $\sim10^{-3}$, while in the EC+SSC models $\epsilon_{Be}$ would be rather larger $\sim0.1$ \cite{Abdo:2010dsc,Boettcher:2013wxa}. Consequently, the EC+SSC models give $\gag\approx 3\times10^{-10}~\mathrm{GeV}^{-1}\sqrt{1~\mathrm{kpc}/r_\mathrm{max}}$, therefore the ALP interpretation in the leptonic models are excluded by CAST. In the hadronic models with the equipartition ratio $\epsilon_{Be}\sim25$ \cite{Boettcher:2013wxa}, we can get $\gag\approx 2\times10^{-11}~\mathrm{GeV}^{-1}\sqrt{1~\mathrm{kpc}/r_\mathrm{max}}$.
\begin{figure}[htbp]
  \includegraphics[width=1.0\columnwidth]{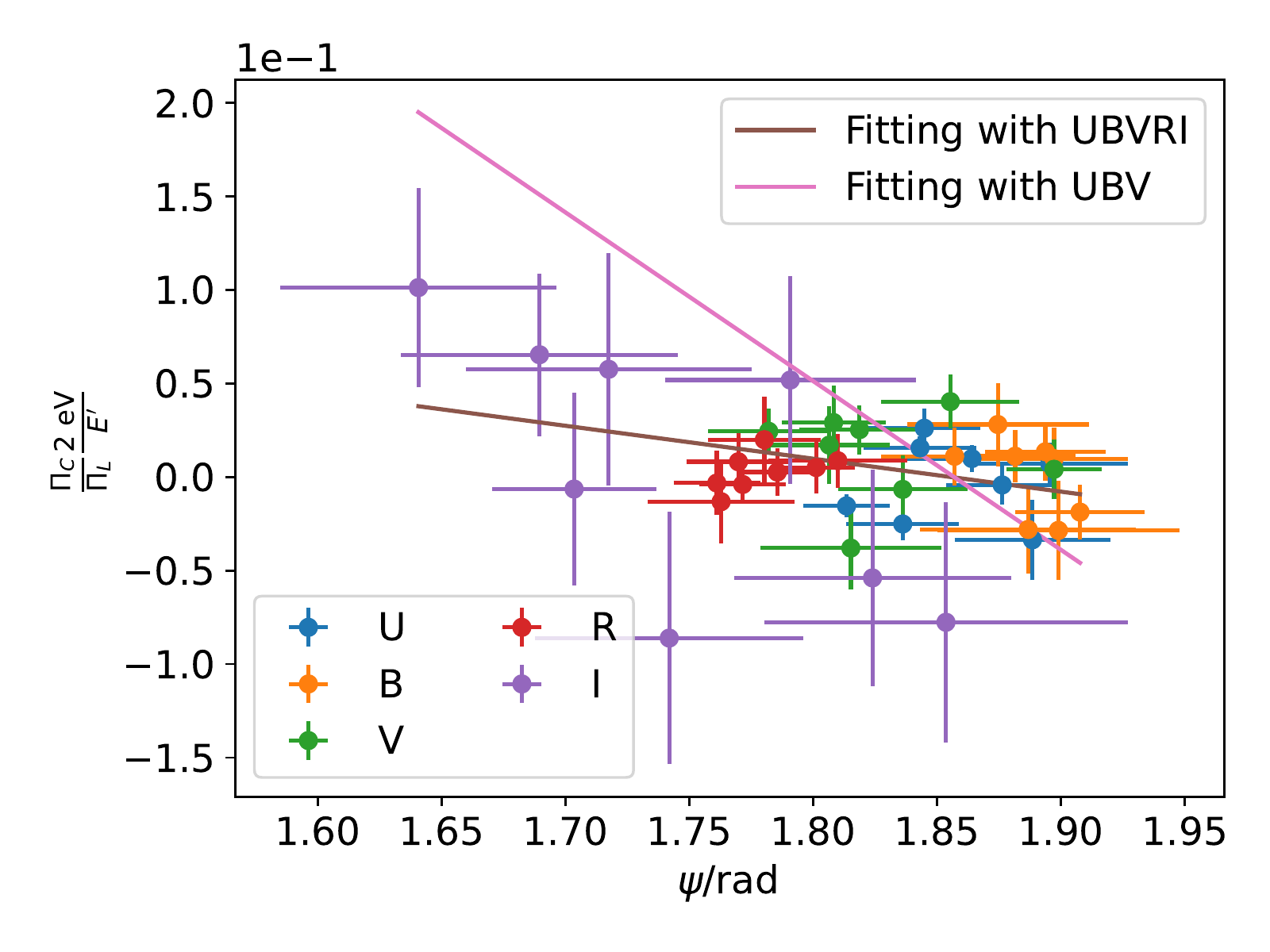}
  \caption{\label{fig:OJ2871993}Linear polynomial fit between the polarization angle $\psi$ and $\frac{\Pi_\mathrm{C}}{\Pi_\mathrm{L}}\frac{2~\mathrm{eV}}{E^\prime}$, using polarization observation data of OJ 287 in \cite{takalo:1993simultaneous}. The letters UBVRI represent different band filters.}
\end{figure}
\begin{table}[htbp]
\begin{tabular}{ccccc}
\hline
\hline
\addlinespace[0.3ex]
~~Bands~~ & ~~$a_0$~~ & ~~$\sigma_{a_0}$~ & ~~$a_1$~ & ~~$\sigma_{a_1}$~~ \\
\addlinespace[0.3ex]
\hline
\addlinespace[0.5ex]
UBVRI & 3.3 & 1.5 & $-1.8$ & 0.79 \\
\addlinespace[0.3ex]
UBV & 16.8 & 6.4 & $-9.0$ & 3.5 \\
\hline
\end{tabular}
\caption{\label{tab:OJ287fit}Coefficients of linear polynomial fit to observation in OJ 287 in form of $\frac{\Pi_\mathrm{C}}{\Pi_\mathrm{L}}\frac{2~\mathrm{eV}}{E^\prime}=a_0+a_1\psi$. Parameters are in unit of $10^{-1}$.}
\end{table}

Apart from 3C 66A, Takalo and Sillanpaa indicated a rapid variability in the U-band CP during the observation of OJ 287. Although the nightly average CP observation did not find optical CP, we can still repeat the above analysis procedure for this object, and provide the results in Fig.~\ref{fig:OJ2871993} and Table.~\ref{tab:OJ287fit}. Similarly, we get a model-dependent estimation for $\gag$, listing in Table.~\ref{tab:OJ287gag}.

The observations of these objects indicate a similar estimation for $\gag$, which is about a few orders of $10^{-11}~\mathrm{GeV}^{-1}$. However, we emphasize that the validity of such results strongly depends on the accuracy of the optical polarimetry. Simultaneous LP and CP measurements in the optical band with high accuracy are needed to verify this analysis in the future.

\begin{table}[htbp]
\begin{tabular}{cccc}
\hline
\hline
\addlinespace[0.3ex]
\multirow{2}{*}{~~Bands~~} & \multirow{2}{*}{$\phi(rad)$} & \multicolumn{2}{c}{~~~$\gag(\times10^{-11}~\mathrm{GeV}^{-1})$~~~} \\
\addlinespace[0.3ex]
\cline{3-4}
\addlinespace[0.5ex]
 &  & ~~~Leptonic~~~ & \multicolumn{1}{l}{~~Hadronic~~~} \\
 \addlinespace[0.3ex]
 \hline
 \addlinespace[0.5ex]
UBVRI & 0.26 or 3.40 & 1.3 & 0.70 \\
\addlinespace[0.3ex]
UBV & 0.30 or 3.44 & 2.9 & 1.6\\
\hline
\end{tabular}
\caption{\label{tab:OJ287gag}Estimation of $\phi$ and $\gag$ using fitting coefficients in OJ 287. The factor $\sqrt{1~\mathrm{kpc}/r_\mathrm{max}}$ in $\gag$ is omitted. $\delta=15$. $f_\gamma\approx1.2,~\epsilon_{Be}=8.2$ for leptonic model and $f_\gamma\approx0.5,~\epsilon_{Be}=66$ for hadronic model \cite{Boettcher:2013wxa}.}
\end{table}
\section{\label{sec:discussion}Discussions}
As mentioned in Sec.~\ref{subsec:result}, Eq.~(\ref{eq:blazarVsol}) is idealized in some sense. In this section, we discuss some relevant physical factors that may affect the above results.
\subsection{Intrinsic CP}
On the astrophysical origin of optical CP in blazars, the inverse Compton scattering of radio photons with high CP and the intrinsic CP are two likely options \cite{rieger:2005possible}. However, the former mechanism requires the significant radio CP and SSC contribution to optical continuum, which are not observed. Therefore, we discuss the consequence that the intrinsic CP is present in this subsection.

The synchrotron radiation of particles generates a small degree of intrinsic CP with $\Pi_\mathrm{C}\propto\nu^{-1/2}$ \cite{legg:1968elliptic,melrose:1971degree}. In a pure electron-positron plasma, the CP component cancels, leaving only the LP component. Assuming an electron-positron-proton plasma with $f$ denoting the fraction of the positron in the jet, the degree of the intrinsic CP can be characterized as \cite{rieger:2005possible}
\iffalse
\begin{equation}
\begin{aligned}
  \Pi_\mathrm{C}&\approx 2.8\times10^{-2} \left(\frac{B}{1~\mathrm{G}}\right)^{1/2}\left(\frac{E^{\prime}}{1~\mathrm{eV}}\right)^{-1/2} \\
  &\times\left(\frac{\Pi_\mathrm{L}}{0.71}\right)\left(\frac{\Gamma}{15}\right)^{3/2}(1-2f),
\end{aligned}
\end{equation}
\fi
\begin{widetext}
\begin{equation}
    \Pi_\mathrm{C}\approx 2.8\times10^{-2} \left(\frac{B}{1~\mathrm{G}}\right)^{1/2}\left(\frac{E^{\prime}}{1~\mathrm{eV}}\right)^{-1/2}\left(\frac{\Pi_\mathrm{L}}{0.71}\right)\left(\frac{\Gamma}{15}\right)^{3/2}(1-2f),
\end{equation}
\end{widetext}
where the Lorentz factor $\Gamma$ can be approximated by $\delta$ in the highly collimated regime. Detecting upper limits of the intrinsic CP of two blazars, Liodakis \textit{et al.} \cite{Liodakis:2021els} claimed the exclusion of high-energy emission models with the high magnetic field strength and low positron fraction, like the hadronic models.

\begin{figure}[htbp]
  \includegraphics[width=1.0\columnwidth]{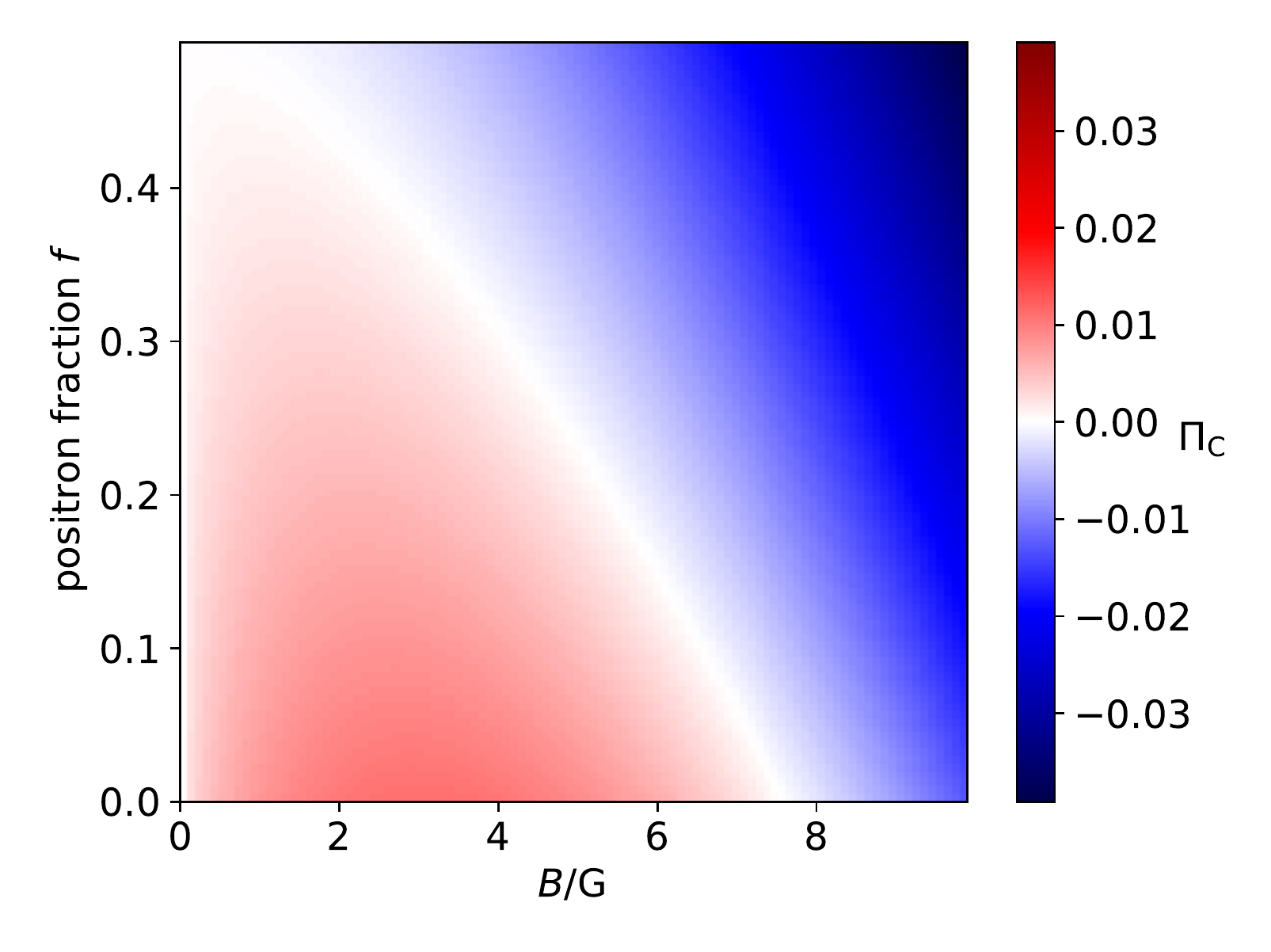}
  \caption{\label{fig:positron} The degree of optical CP of blazars with different magnetic field strength $B$ and plasma composition characterized by $f$.}
\end{figure}

However, the ALP induced CP may offset the intrinsic CP, since the handness of the ALP induced CP can be either positive or negative, which is determined by the angle between the magnetic field and photon polarization. From the view of the observer, the handness of CP is positive, if the direction of the transverse magnetic field lies in $I, III$ quadrants of the Cartesian coordinate system with the direction of polarization as the x-axis. Thus, we study the case that ALPs induce a negative CP in the hadronic model. We calculate the sum of two kinds of CP origins with $B$ in the range of $[0,10]~\mathrm{G}$ and $[0,0.5)$ for $f$. In order to get comparable values, we choose $\gag=5\times10^{-12}~\mathrm{GeV}^{-1}$ and $\phi=\pi/4$ while keeping other parameters unchanged. As shown in Fig.~\ref{fig:positron}, there exists a large parameter region where optical CP does not exceed the precision of measurements even for the large magnetic field strength and low positron fraction. A smaller coupling $\gag$ indicates the possibility of a larger magnetic field. In order to discriminate these two scenarios, it is crucial to measure the radio and optical CP simultaneously. This is because that the radio CP is expected to be associated with the intrinsic CP, while ALPs considered here have little effect on the radio photons.

\subsection{Structure of the jet}\label{subsec:JetStructure}
\begin{figure*}[htbp]
  %\vspace{-0.4cm}
  \subfigure[~$\alpha=0.1,~\mathcal{R}=0.001$]{
  \begin{minipage}[t]{0.3\linewidth}
    \includegraphics[width=\textwidth]{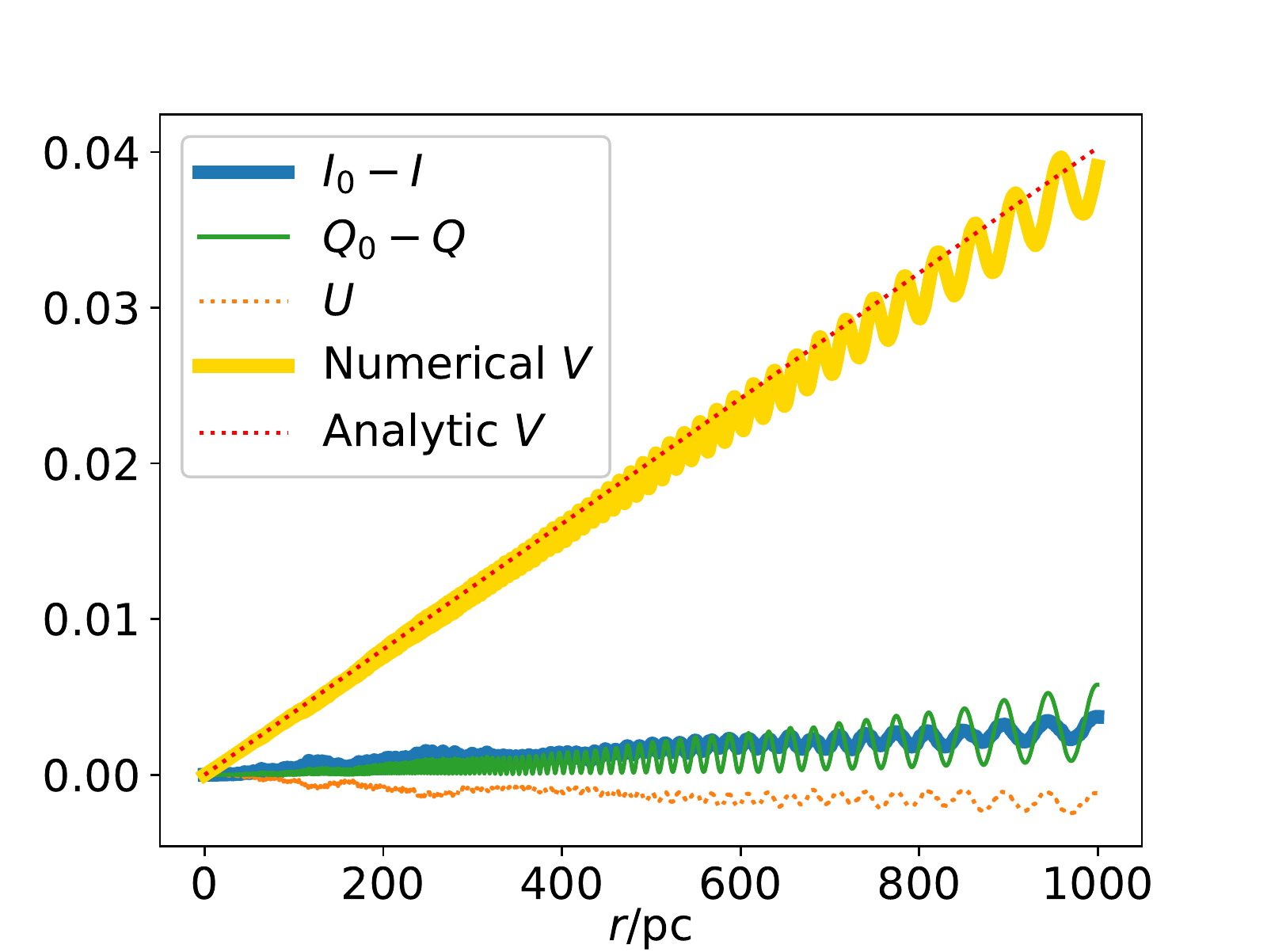}
  \end{minipage}
  }
  \subfigure[~$\alpha=0.2,~\mathcal{R}=0.001$]{
  \begin{minipage}[t]{0.3\linewidth}
    \includegraphics[width=\textwidth]{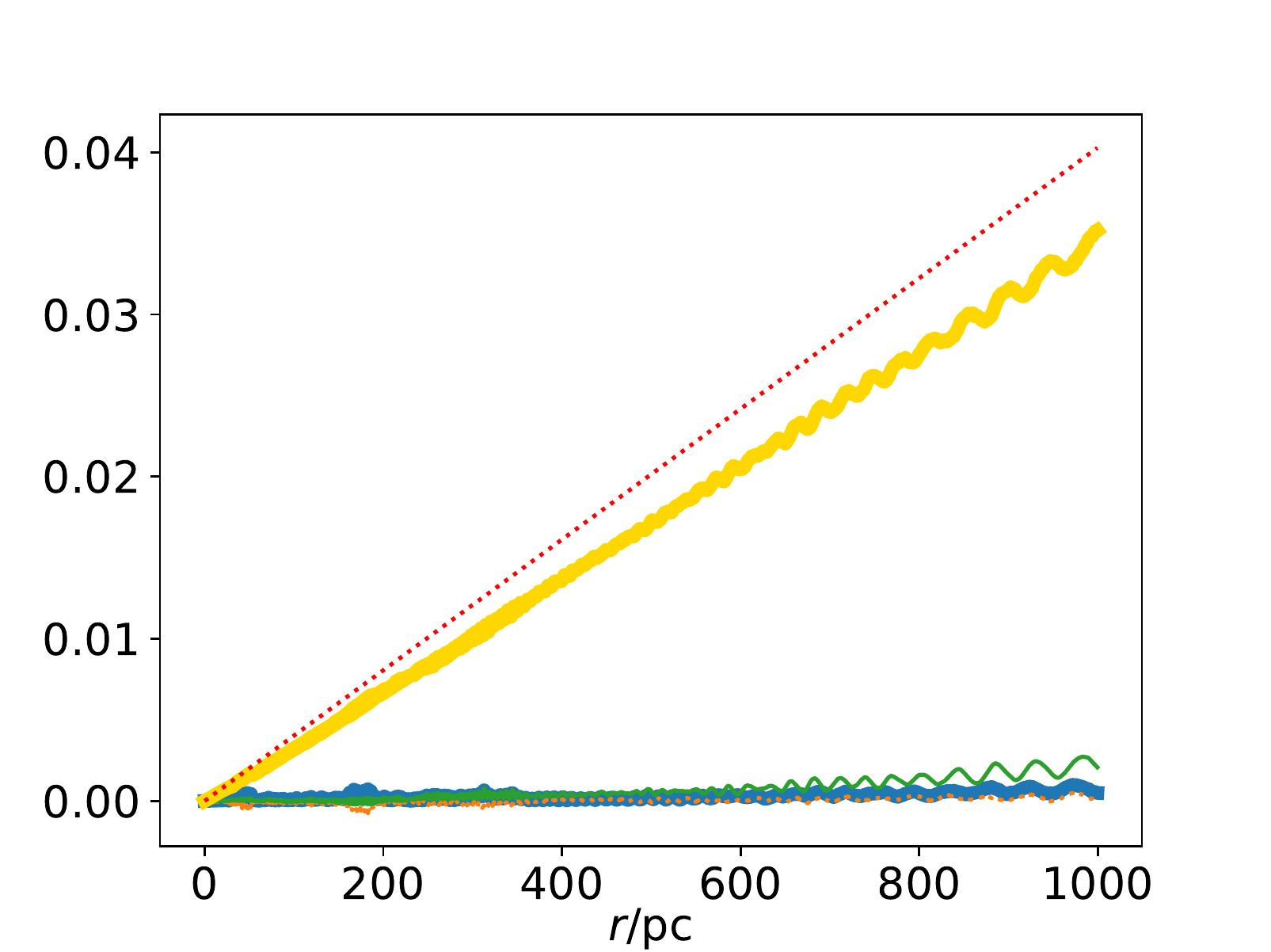}
  \end{minipage}
  }
  \subfigure[~$\alpha=0.3,~\mathcal{R}=0.001$]{
  \begin{minipage}[t]{0.3\linewidth}
    \includegraphics[width=\textwidth]{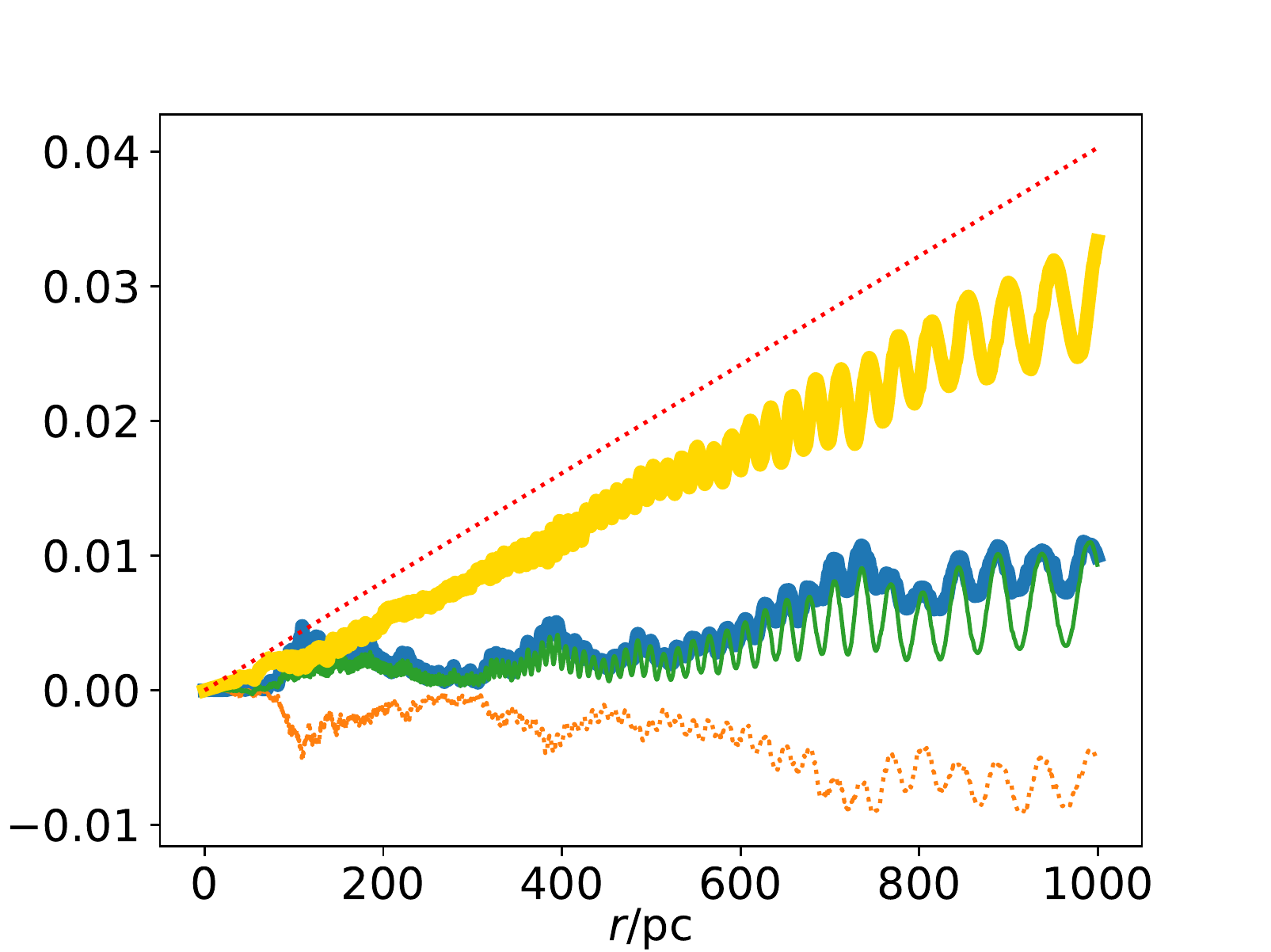}
  \end{minipage}
  }

  \subfigure[~$\alpha=0.1,~\mathcal{R}=0.01$]{
  \begin{minipage}[t]{0.3\linewidth}
    \includegraphics[width=\textwidth]{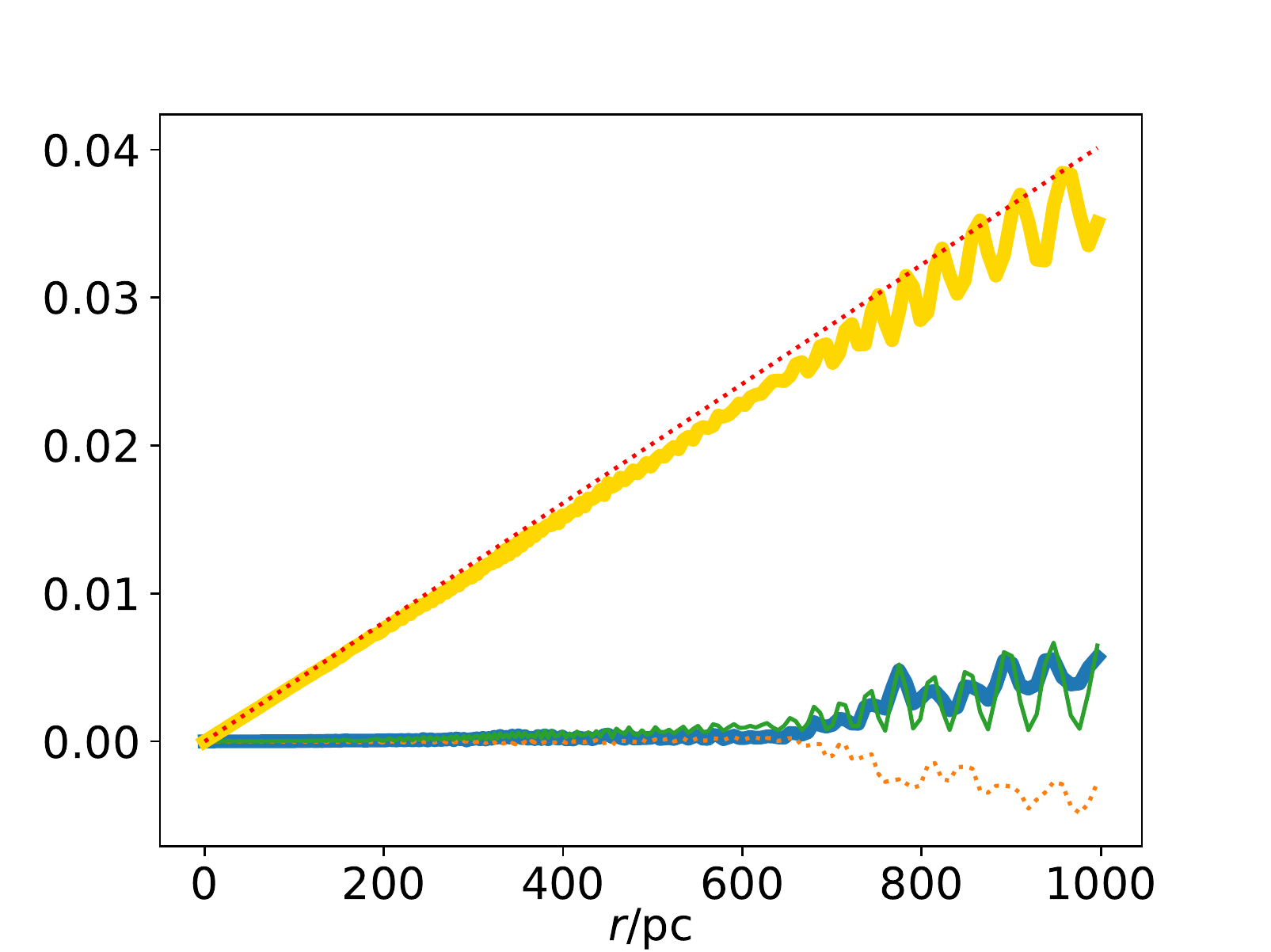}
  \end{minipage}
  }
  \subfigure[~$\alpha=0.2,~\mathcal{R}=0.01$]{
  \begin{minipage}[t]{0.3\linewidth}
    \includegraphics[width=\textwidth]{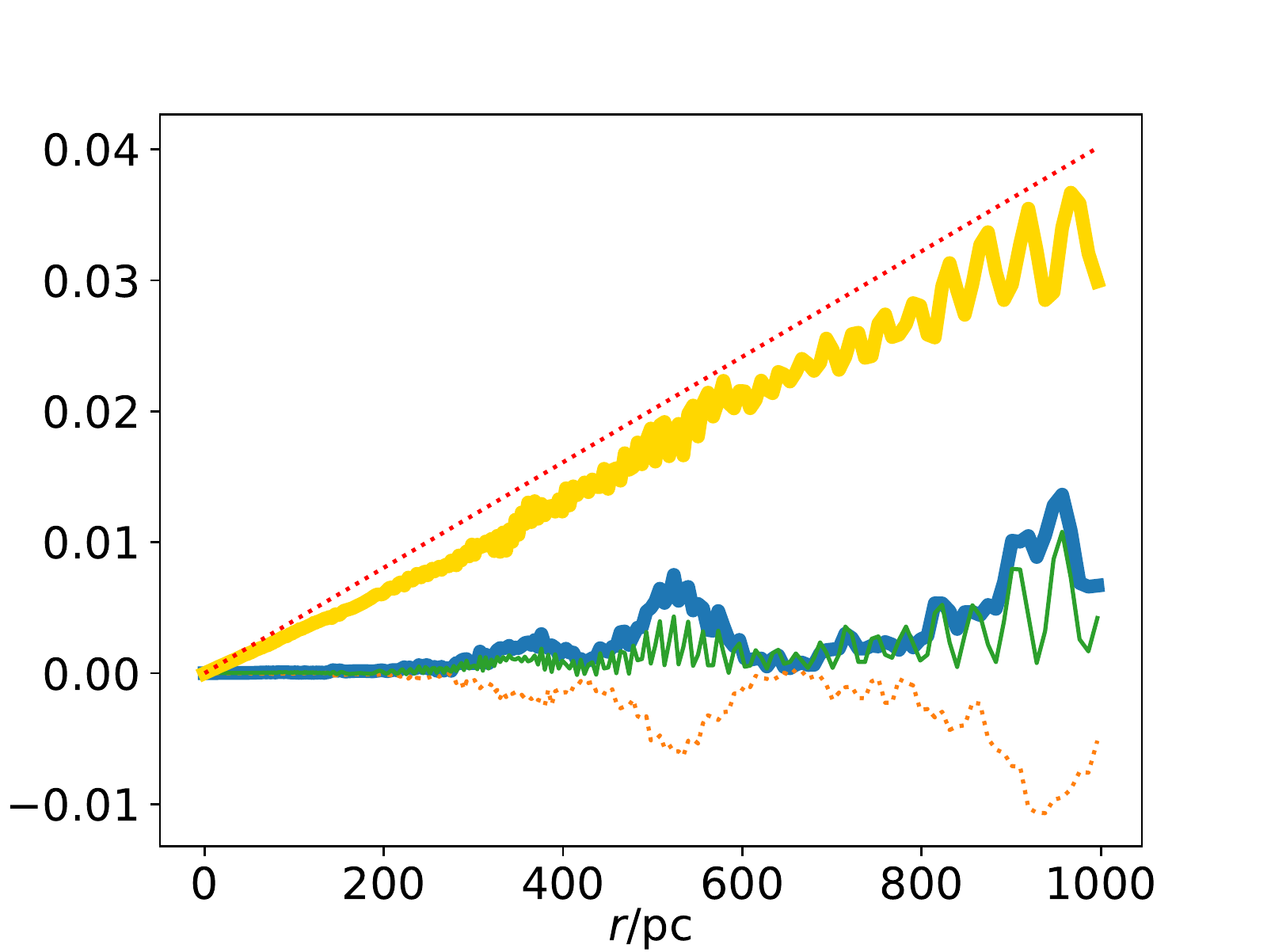}
  \end{minipage}
  }
  \subfigure[~$\alpha=0.3,~\mathcal{R}=0.01$]{
  \begin{minipage}[t]{0.3\linewidth}
    \includegraphics[width=\textwidth]{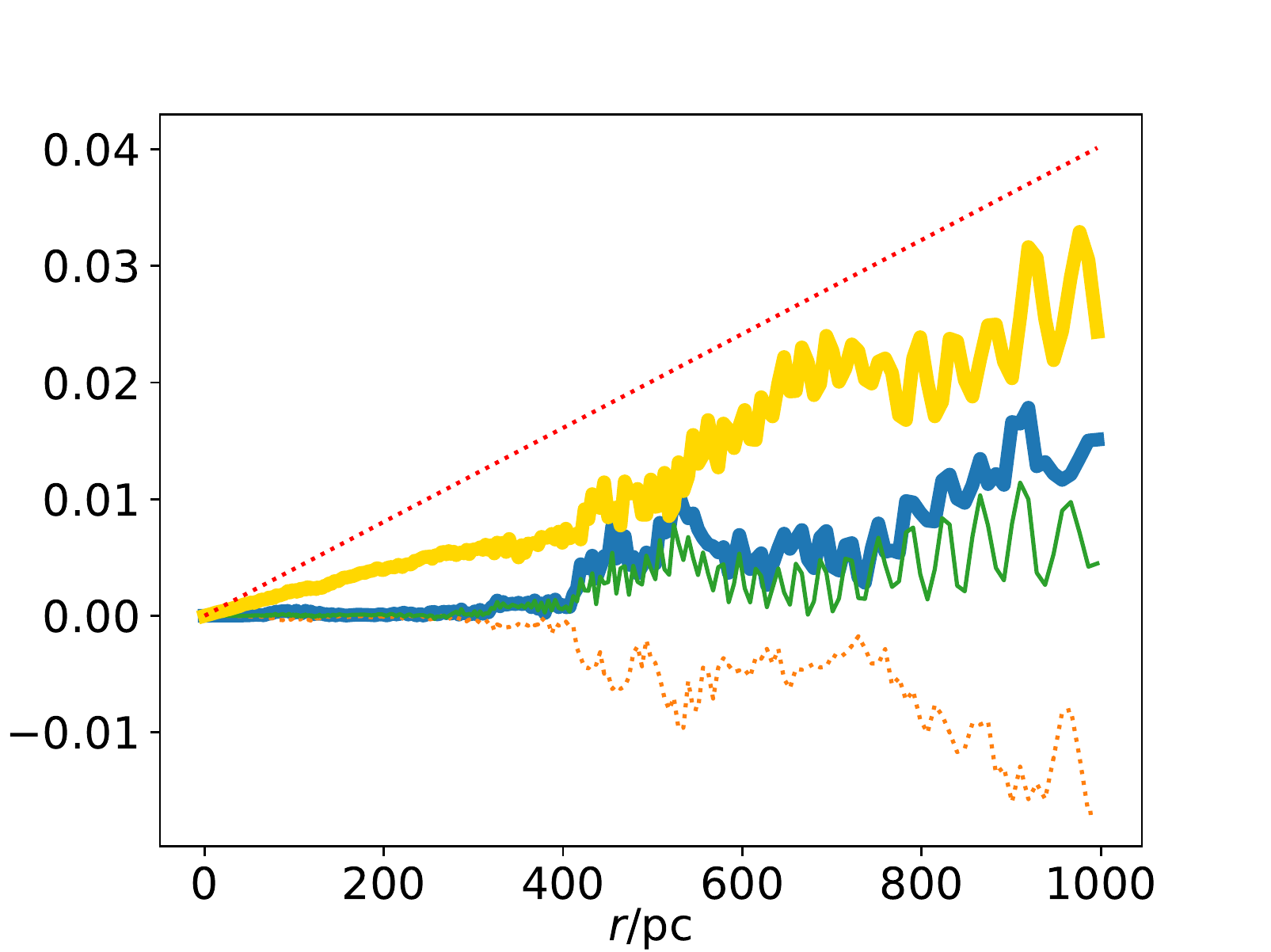}
  \end{minipage}
  }

  \subfigure[~$\alpha=0.1,~\mathcal{R}=0.1$]{
  \begin{minipage}[t]{0.3\linewidth}
    \includegraphics[width=\textwidth]{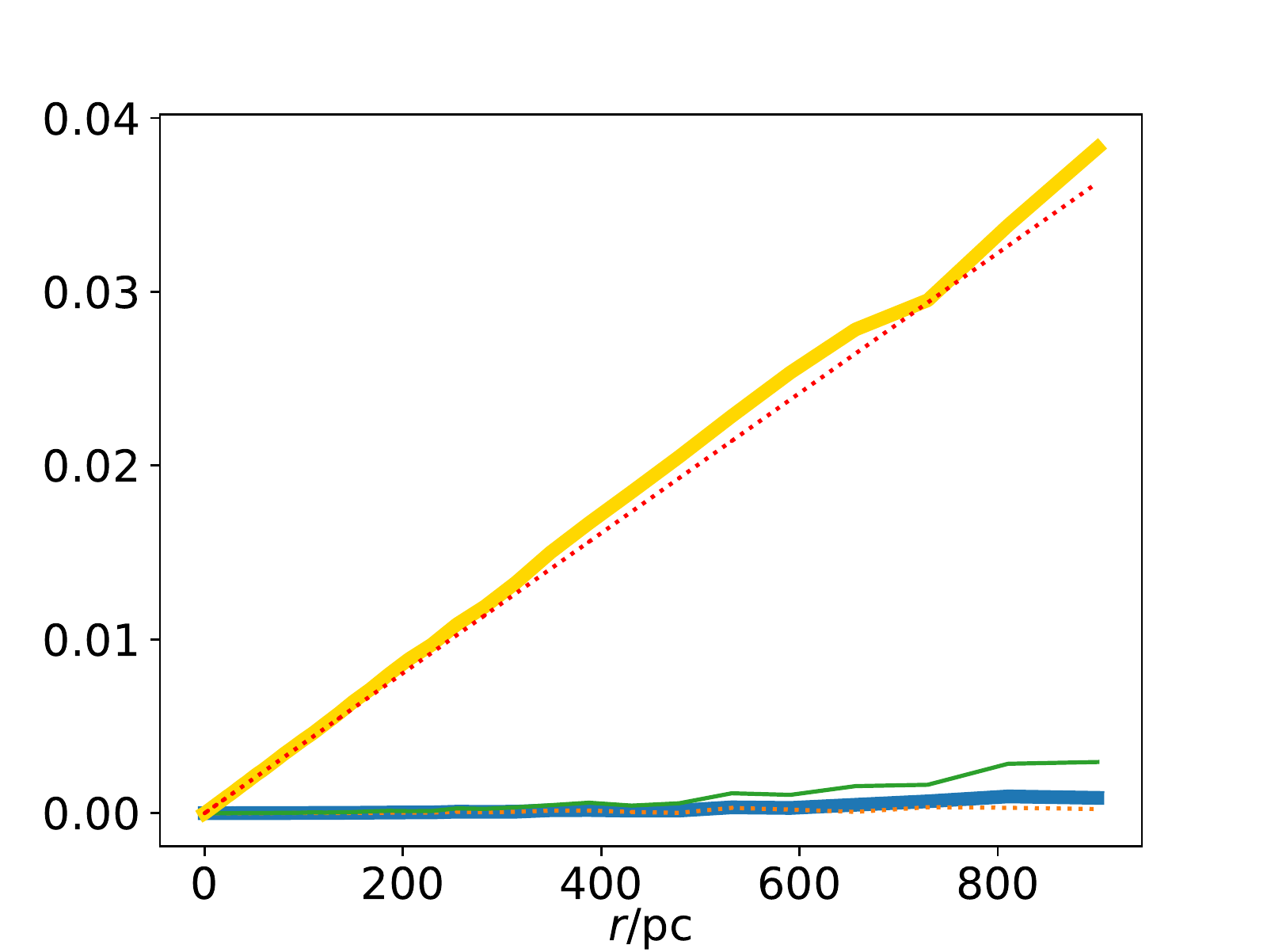}
  \end{minipage}
  }
  \subfigure[~$\alpha=0.2,~\mathcal{R}=0.1$]{
  \begin{minipage}[t]{0.3\linewidth}
    \includegraphics[width=\textwidth]{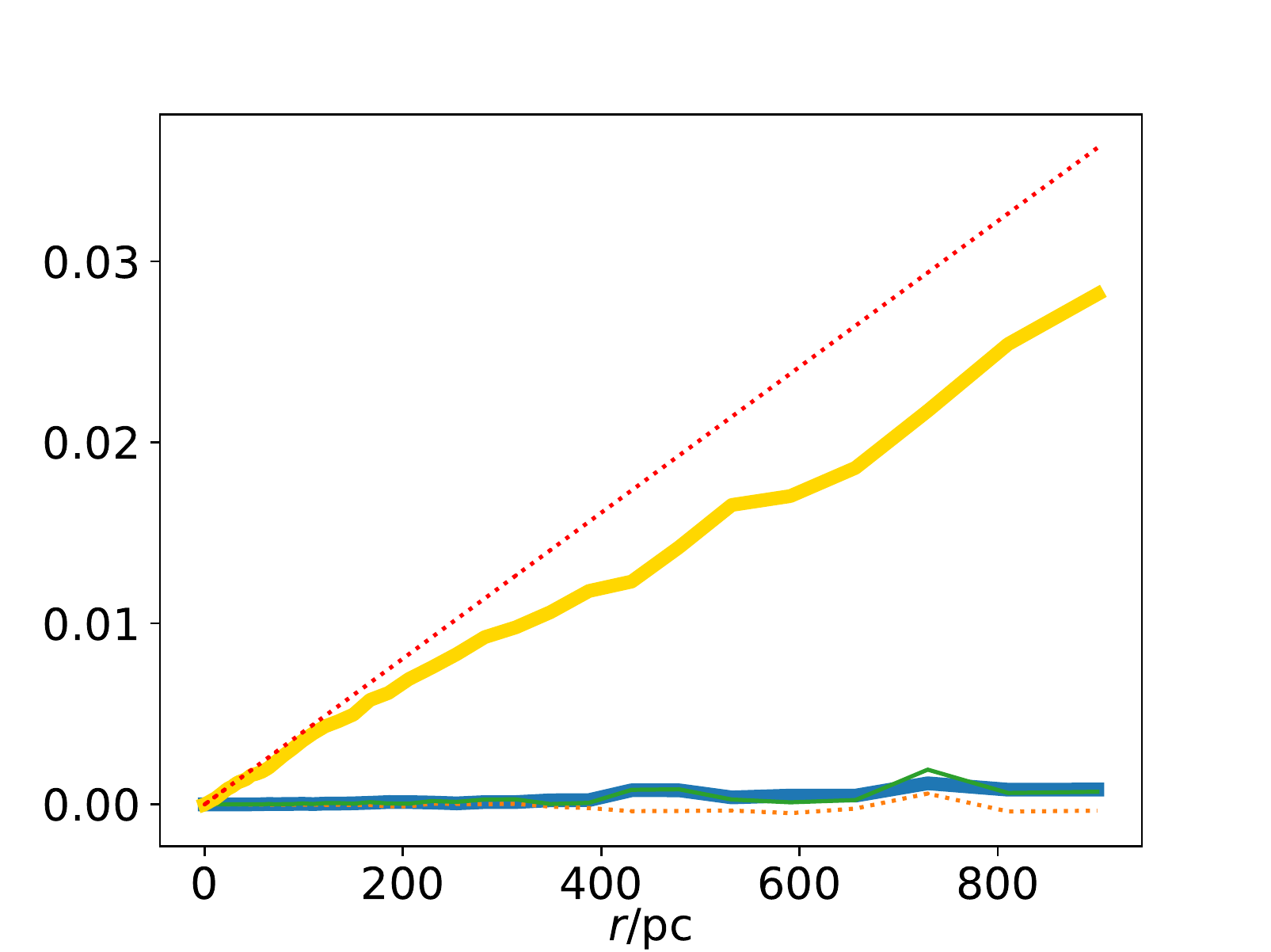}
  \end{minipage}
  }
  \subfigure[~$\alpha=0.3,~\mathcal{R}=0.1$]{
  \begin{minipage}[t]{0.3\linewidth}
    \includegraphics[width=\textwidth]{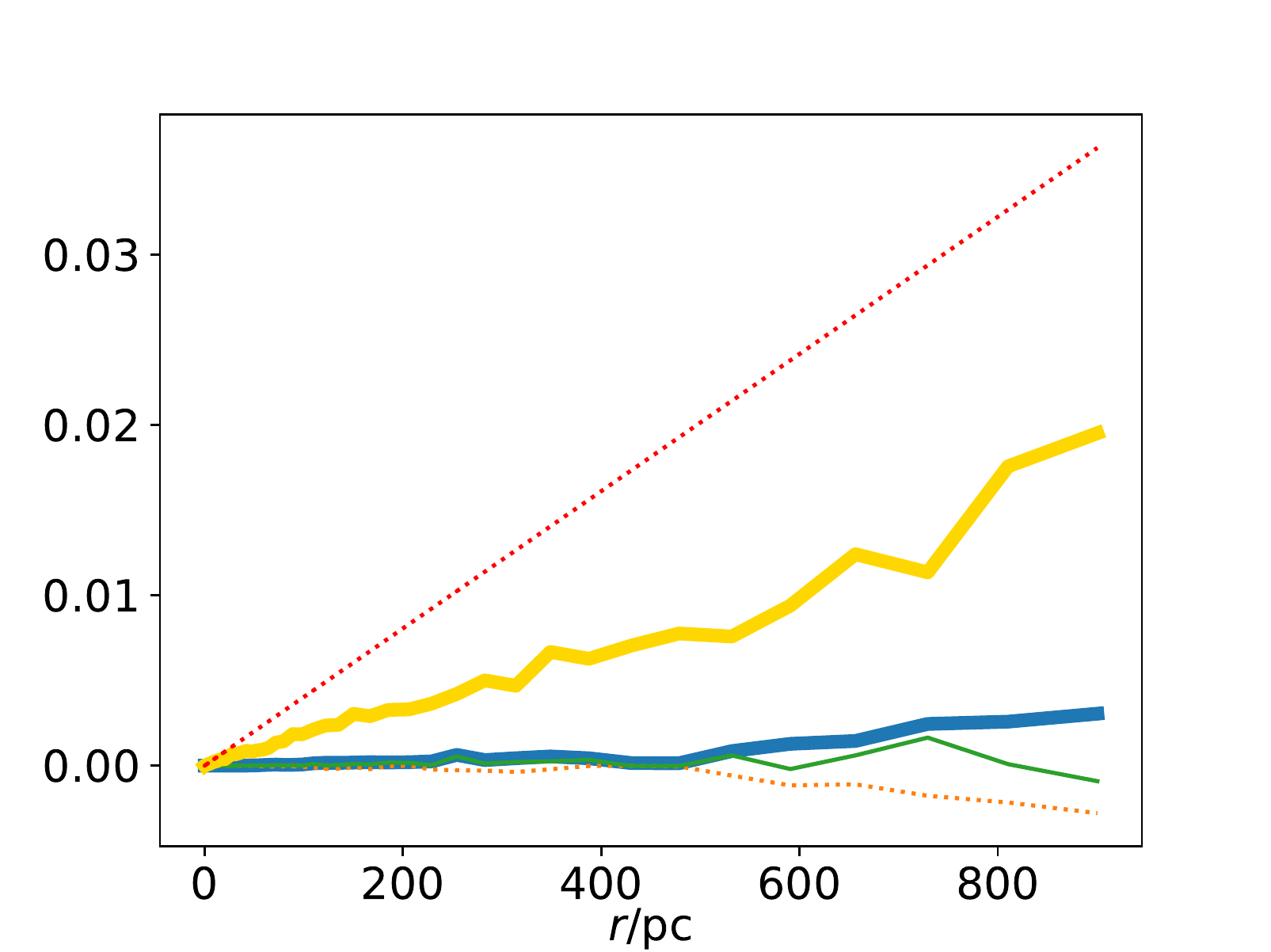}
  \end{minipage}
  }
  \caption{Examples of evolution of Stokes parameters for magnetic field with varied angle in different cases. The configuration of jet is sliced into domains, determined by the decrease ratio $\mathcal{R}$. The angle in each domain is $\phi = \phi_{0} + \alpha\Delta\phi$ with $\phi_{0}=3\pi/4$ and a uniform random variable $\Delta\phi$ in the $[-\pi,\pi]$ range.}
  \label{fig:VariedField}
\end{figure*}

\begin{figure*}[htbp]
  \subfigure[~$\alpha=0.1$]{
  \begin{minipage}[t]{0.3\linewidth}
    \includegraphics[width=\textwidth]{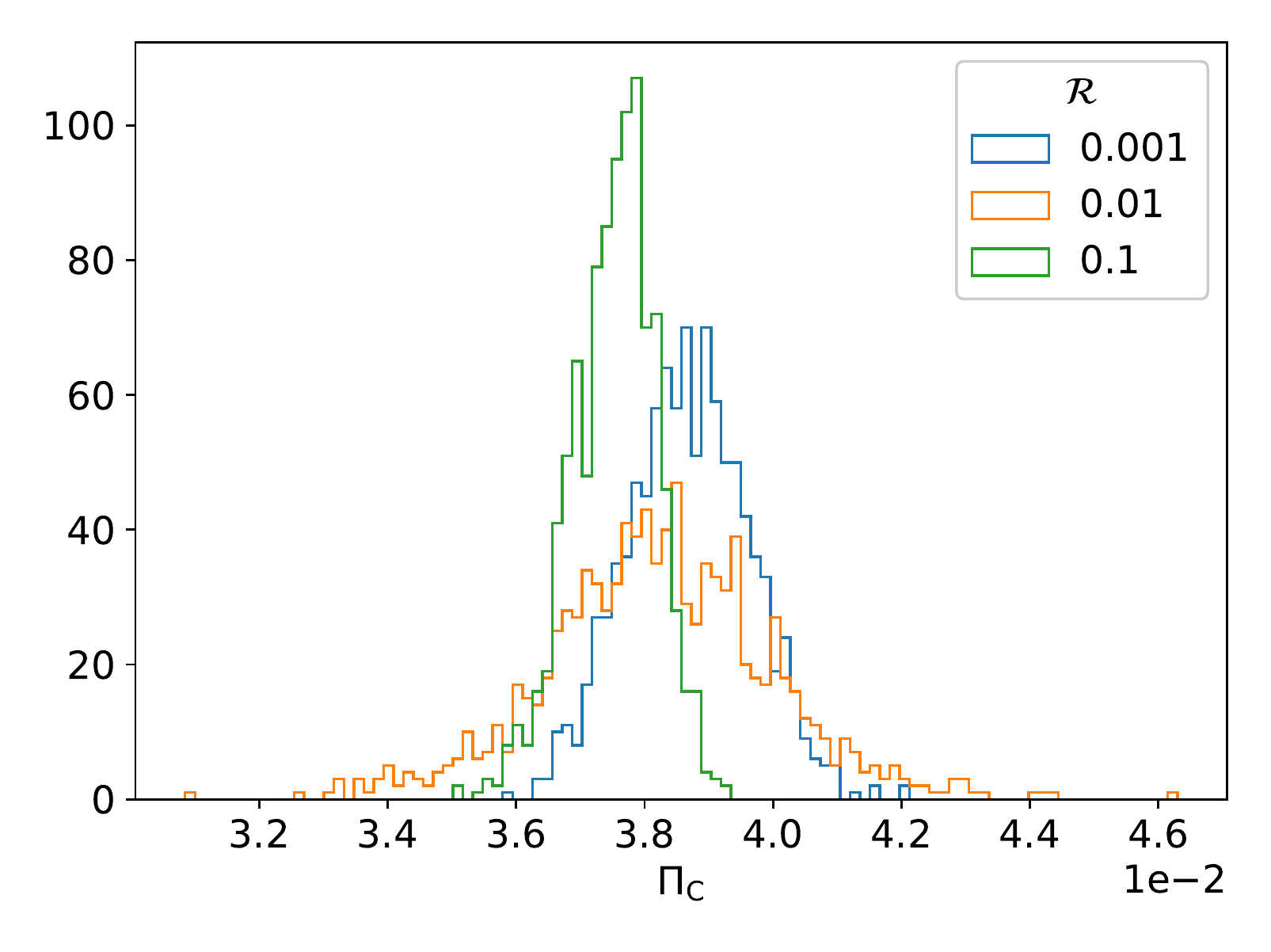}
  \end{minipage}
  }
  \subfigure[~$\alpha=0.2$]{
  \begin{minipage}[t]{0.3\linewidth}
    \includegraphics[width=\textwidth]{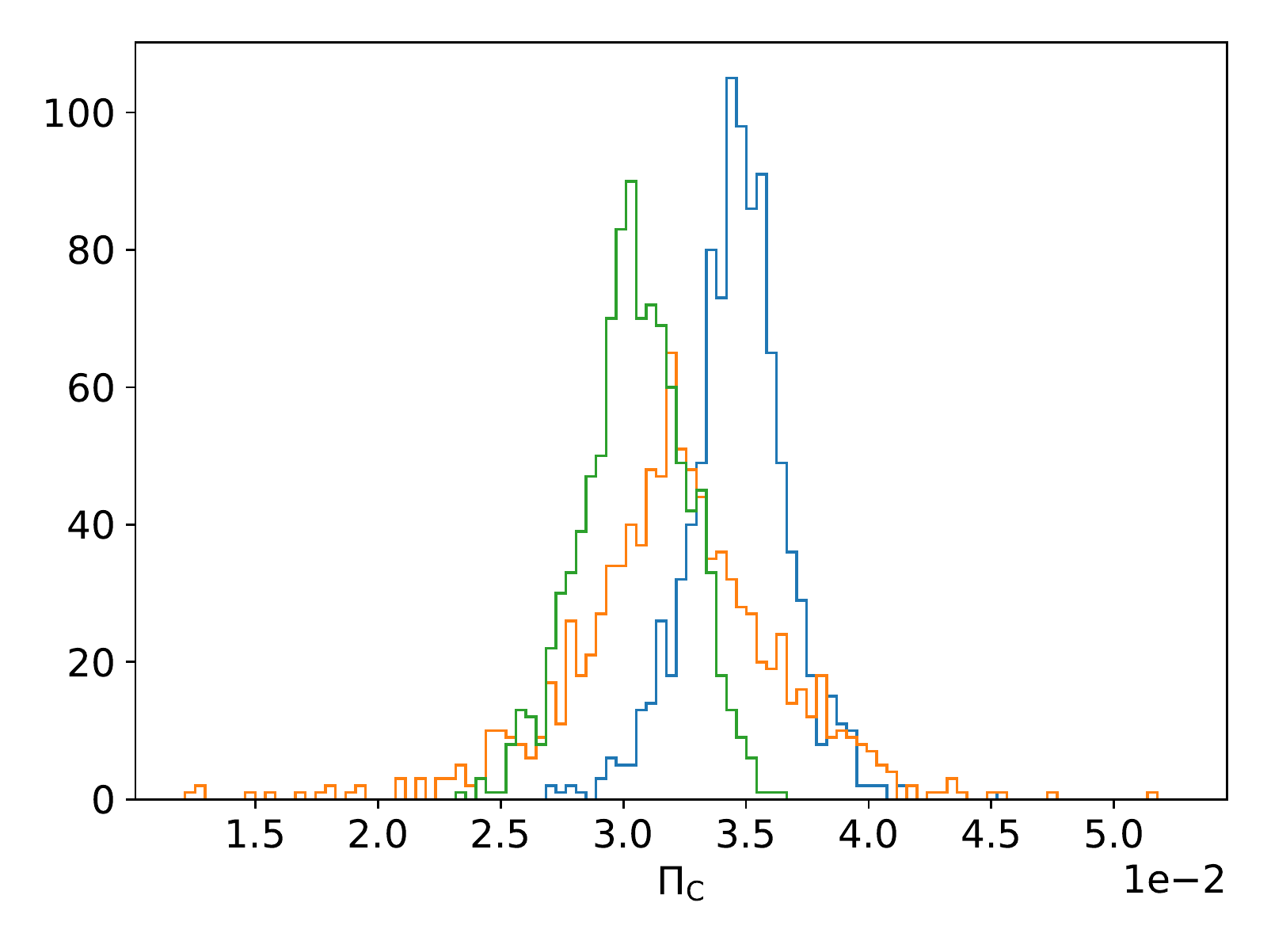}
  \end{minipage}
  }
  \subfigure[~$\alpha=0.3$]{
  \begin{minipage}[t]{0.3\linewidth}
    \includegraphics[width=\textwidth]{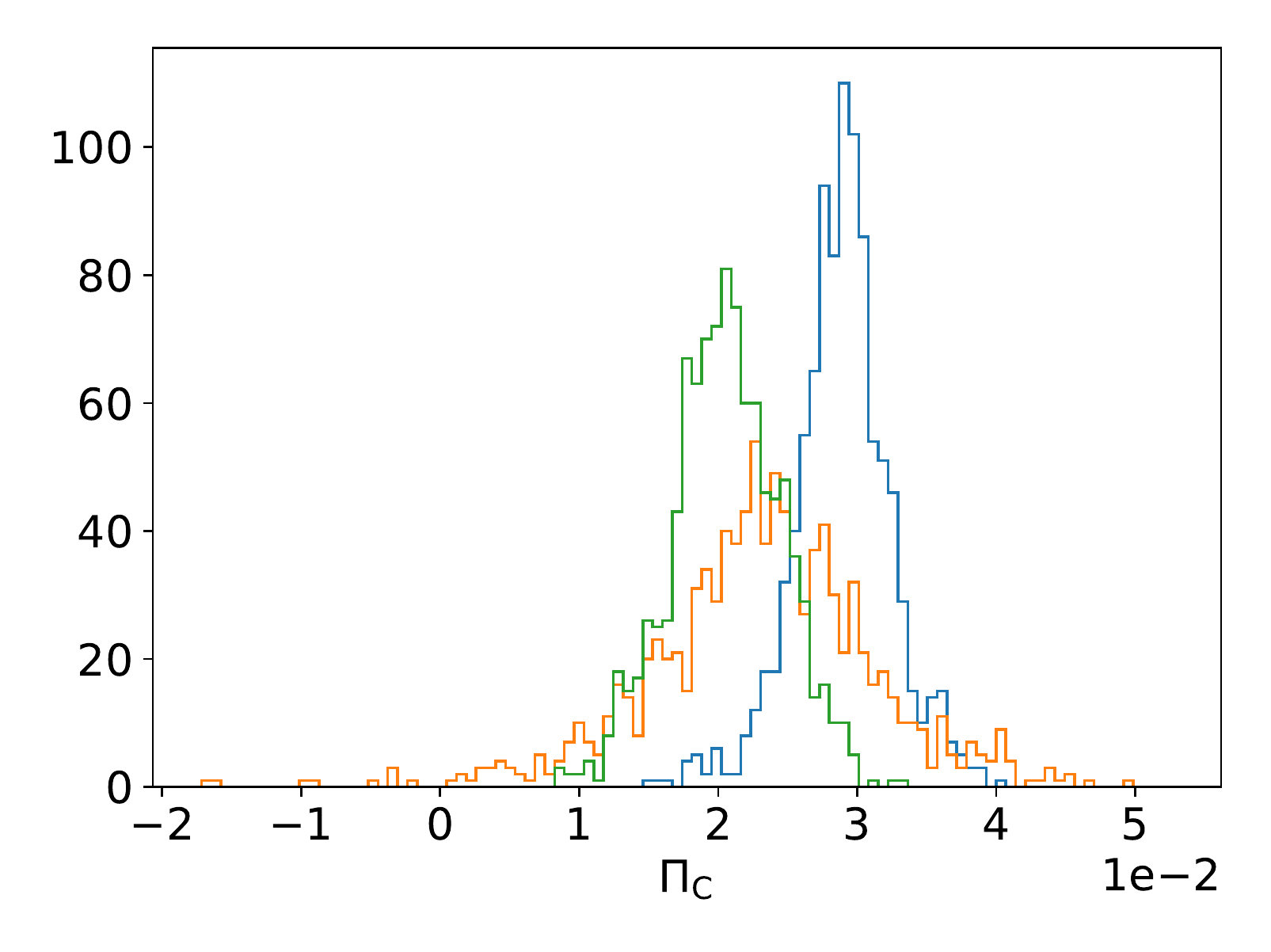}
  \end{minipage}
  }
  \caption{Similar to Fig.~\ref{fig:VariedField}. Distribution of optical CP induced by ALPs for magnetic field with varied angle in different cases.}
  \label{fig:DistributionVariedField}
\end{figure*}

In the above analysis, the angle of the transverse magnetic field is idealized, fixing to $3\pi/4$ or $\pi/4$. However, for a more realistic case, the direction of magnetic field could change  in the jet and result in a shorter coherent length, which in turn leads to a smaller CP. In order to estimate this effect, we assume that the direction of the magnetic field is partially random in each calculation domain, i.e.,
\begin{equation}
  \phi = \phi_{0} + \alpha\Delta\phi,
\label{variation}
\end{equation}
where $\phi_{0}=3\pi/4$ and $\Delta\phi$ is a uniform random variable in the range of $[-\pi,\pi]$. The factor $\alpha$ characterizes the degree of randomness. Similar to what we have done in Sec.~\ref{subsec:result}, examples of the evolution of the Stokes parameters with $\alpha=0.1,0.2,0.3$ and $\mathcal{R}=0.001,0.01,0.1$ are shown in Fig.~\ref{fig:VariedField}. The coherent length of the magnetic field (also the number of domains) is determined by the decrease ratio $\mathcal{R}$. As shown in Fig.~\ref{fig:VariedField}, the values of $\alpha$ and $\mathcal{R}$ can significantly change the pattern of the evolution. As $\alpha$ increases,
%smaller the coherent length is, more important role it plays, and
the evolution shows more randomness and quiver. Moreover, for each set of parameters, we demonstrate the distribution of CP sampled for 1000 sets of magnetic field configurations in Fig.~\ref{fig:DistributionVariedField}. It shows that for a more realistic field configuration, the values of CP are smaller than the optimal case, while its magnitude remains in the same order.

Next, we give a qualitative analysis of the results in Fig.~\ref{fig:VariedField} and Fig.~\ref{fig:DistributionVariedField}. Given the approximation that $\alpha\ll1$ is true, we can rewrite Eq.~(\ref{eq:MathcalV1}) as
\begin{equation}
  \mathcal{V}_\mathrm{n+1}\approx\Pi_\mathrm{L}\sin2(\phi_\mathrm{0}-\psi_\mathrm{n})+2\alpha\Delta\phi_\mathrm{n}\Pi_\mathrm{L}\cos2(\phi_\mathrm{0}-\psi_\mathrm{n}),
\end{equation}
where the subscription $n$ indicates the $n$th domain. Then the change of $V$ is determined by two parts: the incremental part (the first term) and the random part (the second term). Also, Eq.~(\ref{eq:Vsol}) is applicable for the incremental part in each domain, which explains the magnitude of the expectation value of the CP. When $\phi_\mathrm{0}=3\pi/4$, the second term can be written as $-2\alpha\Delta\phi_\mathrm{n}U_\mathrm{n}$. Hence, as shown in Fig.~\ref{fig:VariedField}, the quiver of $V$ in each domain is related to the magnitude of $U$ of the former domain. As for the contribution of the random part, it can be treated as a random walk, thus its variance is estimated to be $2\alpha\langle\Delta\phi_\mathrm{n}U_\mathrm{n}\rangle\sqrt{N}$, where $N$ is the number of domains and the angle bracket means the expectation value. Therefore, with the increase of $\alpha$, the randomness of the evolution of CP degree becomes stronger.

In addition, considering the limit of the spatial resolution of the detector, the photons received by the detector are not from a certain point on the source but from the whole emitting region. In this case, the direction of the transverse magnetic field may be spatial dependent. In order to have a rough idea of this effect, we can replace the $\mathcal{V}$ in Eq.~(\ref{eq:Vsol}) with $\int_{\phi_{1}}^{\phi_{2}}\mathcal{V}(\phi)d\phi/(\phi_{2}-\phi_{1})$, assuming the direction angle of the transverse magnetic field in the region evenly varies from $\phi_{1}$ to $\phi_{2}$ and the initial photons from different positions have the same polarization state. In practice, the density matrices of photons are calculated in multi paths, where Eq.~(\ref{variation}) is still adopted but the values of $\phi_{0}$ evenly vary in different paths. Then summing all density matrices, we obtain the assembly CP in a specific range of $\phi_{0}$ in the numerical calculation.

\begin{figure*}[htbp]
  \subfigure[~$\phi_\mathrm{0}:0\sim2\pi$]{
  \begin{minipage}[t]{0.3\linewidth}
    \includegraphics[width=\textwidth]{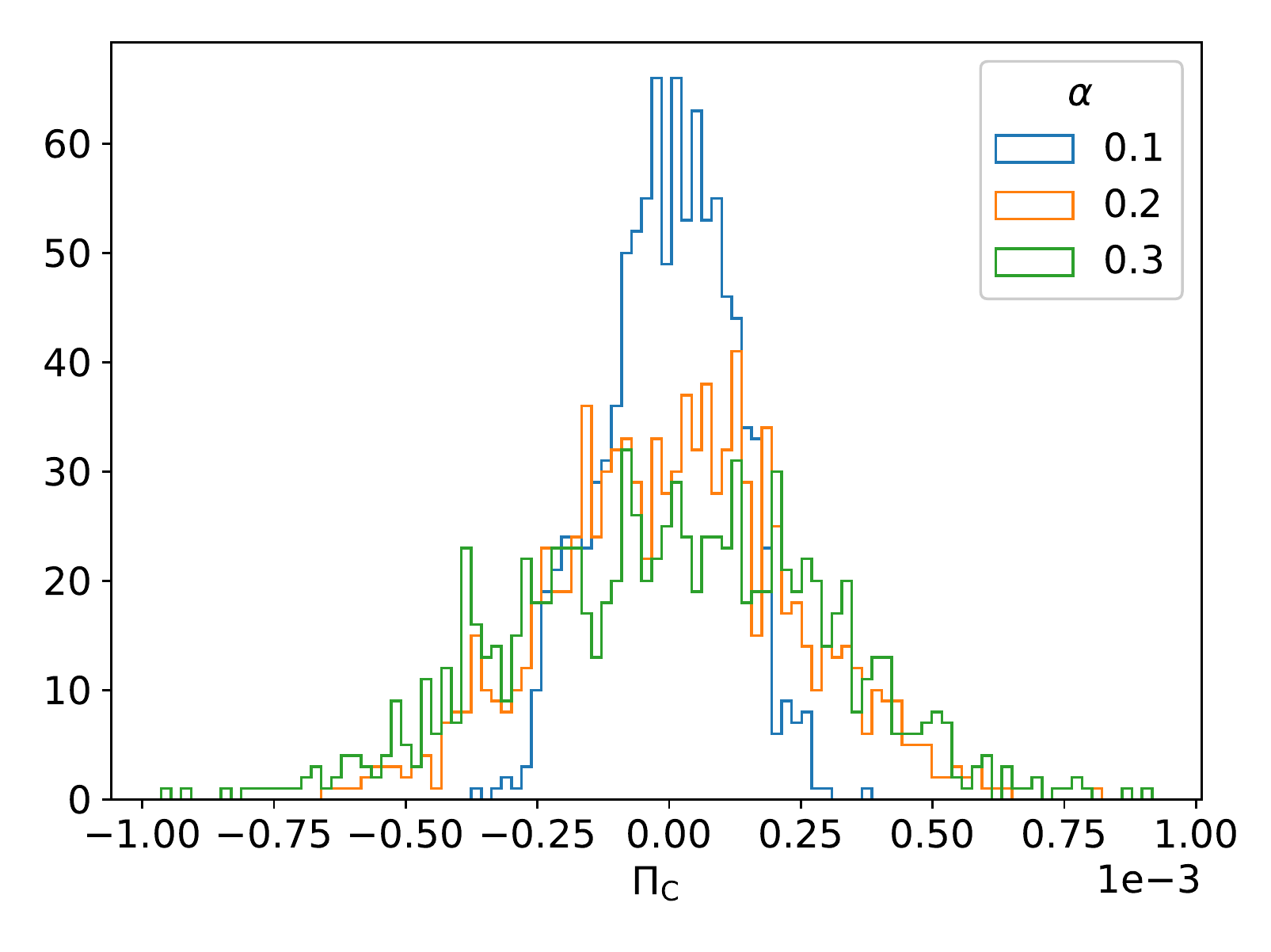}
  \end{minipage}
  }
  \subfigure[~$\phi_\mathrm{0}:0\sim\pi$]{
  \begin{minipage}[t]{0.3\linewidth}
    \includegraphics[width=\textwidth]{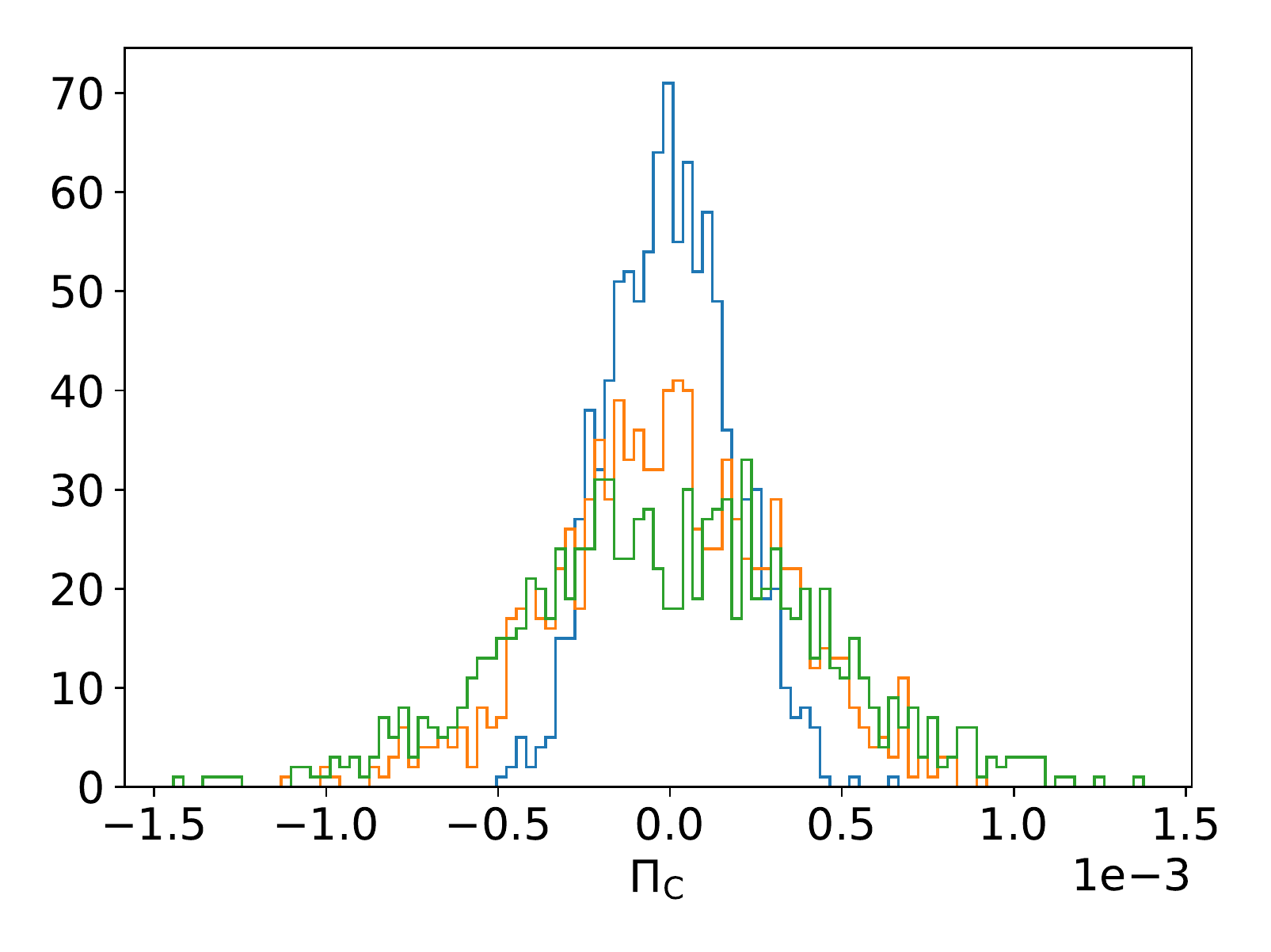}
  \end{minipage}
  }
  \subfigure[~$\phi_\mathrm{0}:0\sim\pi/2$]{
  \begin{minipage}[t]{0.3\linewidth}
    \includegraphics[width=\textwidth]{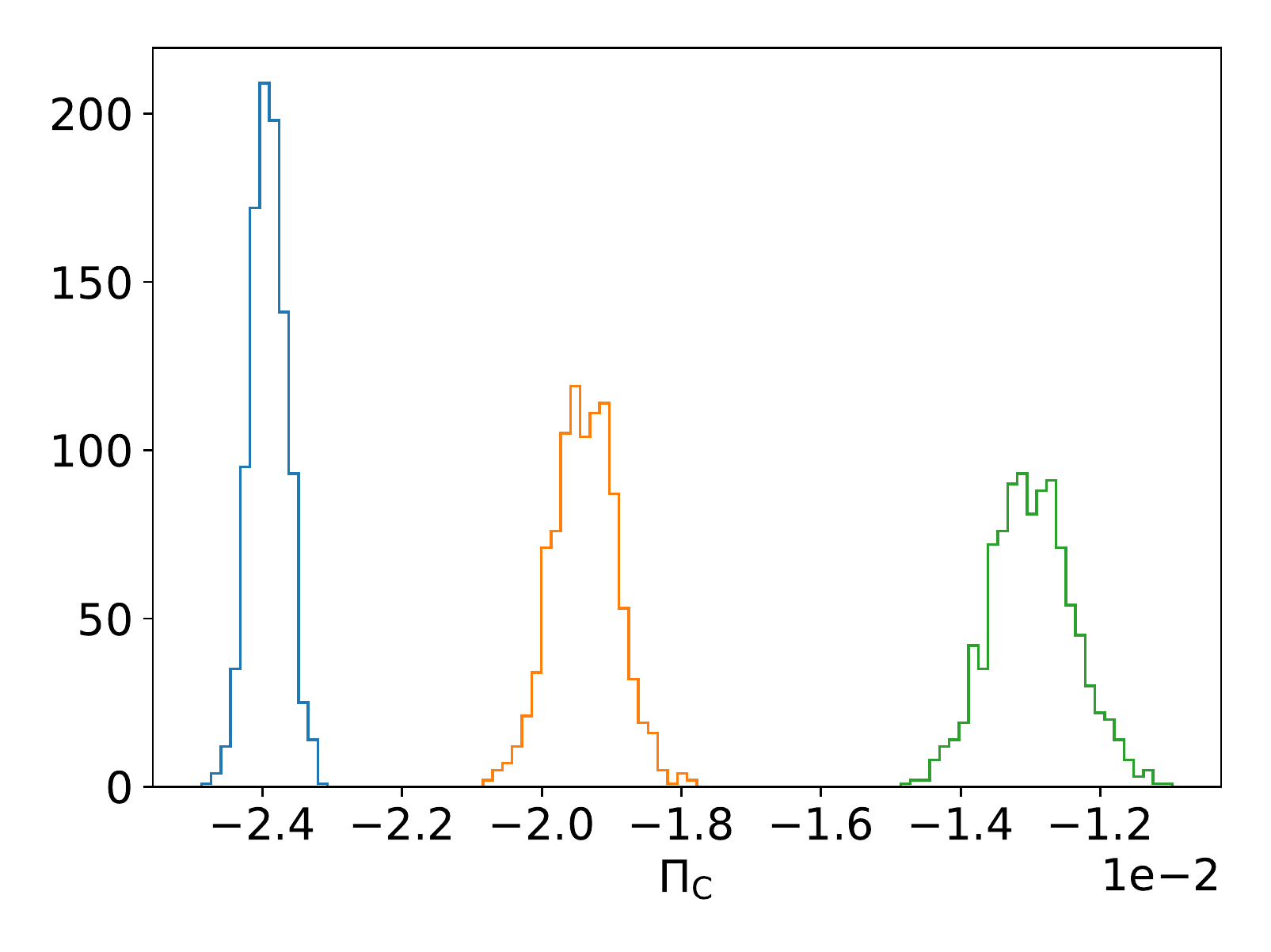}
  \end{minipage}
  }
  \caption{Distribution of assembly optical CP for partially random magnetic field with $\phi_{0}$ in different intervals. The decrease ratio is chosen to be $\mathcal{R}=0.1$ for simplicity.}
  \label{fig:DistributionExtendSource}
\end{figure*}

In Fig.~\ref{fig:DistributionExtendSource}, we show the distribution of assembly CP for the partially random magnetic fields in three intervals of $\phi_{0}$. Note that a toroidal transverse magnetic field for $\phi_{0}$ is assumed here. This indicates that the transverse magnetic field roughly follows a uniform distribution. The decrease ratio is chosen to be $\mathcal{R}=0.1$ for simplicity.
These subfigures can be recognized as the distribution of the observed CP in blazars with different measurement precision. We find that the assembly CP significantly cancels in the first two cases, leading to a zero expectation, while the third case shows a moderate CP. This shows that high-precision measurements are needed to observe optical CP in a large sample. Nevertheless, even in the low-precision cases, it is possible that there are some blazars that can produce observable CP.

Note that the above results base on the assumption that the transverse magnetic field roughly follows a uniform distribution.
However, the most probable distribution of the direction of the magnetic field is very likely to significantly deviate from the uniform distribution in reality, considering the relativistic beaming effect, the jet angle with respect to the line-of-sight, and the complexity of the structure of the jet magnetic field. Consequently, the expectation of the realistic observed CP distribution is unlikely to be zero and would still indicate possible significant signatures. However, a complete and accurate survey of the distribution depends on an in-depth study of the magnetic field configuration in blazars. The detailed study on this issue is beyond the scope of this paper, so it is left to follow-up research. More studies on the blazar jets and high-precision measurements are needed to get more reliable conclusions.
%\begin{widetext}
%\end{widetext}
\subsection{CP in other astrophysical magnetic fields}
Magnetic fields are ubiquitous in the propagation of photons from blazars to Earth. The ALP-photon mixing in other astrophysical magnetic fields may contribute to optical CP and alter the results obtained above. For completeness, we give a rough estimation of these effects. According to Eq.~(\ref{eq:EstimateV}), the contribution of mixing is $\sim\Pi_\mathrm{L}\gag^{2}B^{2}l_\mathrm{osc}L$ as long as the weak-mixing condition is fulfilled. Here we define a new quantity, i.e.,
\begin{equation}
    f_\mathrm{CP}=\left(\frac{B}{1~\mathrm{G}}\right)^{2}\left(\frac{n_\mathrm{e}}{5\times10^{4}~\mathrm{cm}^{-3}}\right)^{-1}\left(\frac{L}{1~\mathrm{kpc}}\right),
\end{equation}
which straightforwardly characterizes the contribution to CP due to the ALP-photon mixing in different magnetic fields. For example, for BL Lacs considered in this paper, we can get $f_\mathrm{CP}\simeq1$.
In Table.~\ref{tab:fCP}, we show the value of $f_\mathrm{CP}$ for different magnetic field scenarios, including the magnetic field inside a galaxy cluster (intra-cluster magnetic field, ICMF), the intergalactic magnetic field (IGMF), and the galactic magnetic field in the Milky Way (GMF). For each scenario, the typical values of $B$, $n_\mathrm{e}$, and coherent length scale $L$ are also given. Considering that the turbulent magnetic field configuration is usually used in these fields, the situation is closer to the random walk case that we discuss in the former part, and the output relies on the specific realization of the magnetic field. From Table.~\ref{tab:fCP}, we can see that the influence of other astrophysical magnetic fields can be neglected.
\begin{table}[htbp]
\begin{tabular}{ccccc}
\hline
\hline
\addlinespace[0.3ex]
~~~~~~Scenarios~~~~~~ & ~$B(\mathrm{G})$~ & ~$n_\mathrm{e}(\mathrm{cm}^{-3})$~ & ~$L(\mathrm{kpc})$~ & ~$f_\mathrm{CP}$~ \\
\addlinespace[0.3ex]
\hline
\addlinespace[0.5ex]
ICMF \cite{Govoni:2004as,Feretti:2012vk} & $10^{-6}$ & $10^{-3}$ & $10$ & $5\times10^{-4}$ \\
\addlinespace[0.3ex]
IGMF \cite{Blasi:1999hu,DeAngelis:2007rw,PierreAuger:2010ofq,WMAP:2012nax} & $10^{-9}$ & $10^{-7}$ & $5\times10^{4}$ & $2.5\times10^{-2}$ \\
\addlinespace[0.3ex]
%SCMF \cite{Kravtsov:2001ac,Vall_e_2002,Xu:2005rb,Vallee:2011zz} & $10^{-6}$ & $10^{-6}$ & $10^{2}$ & $5$ \\
GMF \cite{Jansson:2012pc,Jansson:2012rt} & $10^{-6}$ & $10^{-1}$ & $10^{-2}$ & $5\times10^{-9}$ \\
\hline
\end{tabular}
\caption{\label{tab:fCP}The value of $f_\mathrm{CP}$ for different magnetic field scenarios.}
\end{table}

Apart from the astrophysical magnetic fields mentioned above, the field in the supercluster has been discussed in \cite{Payez:2011sh,Payez:2012vf}. The ALP-photon mixing in the supercluster was originally proposed to explain the alignments of the optical polarization of light from quasars in extremely large regions of the sky, but disfavored by subsequent study. The effect of the supercluster needs to be taken into account when it intervenes in the line-of-sight of blazars. The authors of \cite{Payez:2012vf} claimed to constrain $\gag\lesssim10^{-13} ~\mathrm{GeV}^{-1}$ for $m_\mathrm{a}\lesssim10^{-14}~\mathrm{eV}$, which would be the most stringent limit to date. They assumed the electron density of $\mathcal{O}(10^{-6})~\mathrm{cm}^{-3}$ and the magnetic field of $\mathcal{O}(\mu\mathrm{G})$ with the coherent length $\sim100~\mathrm{kpc}$ in the supercluster. This configuration of the supercluster field would give $f_\mathrm{CP}\sim5$. However, the studies of supercluster magnetic fields \cite{Dolag:2004kp,Bregman:2004xa,Suhhonenko:2011mh,Einasto:2014hda,Tully:2014gfa,Einasto:2020yzp} find that their structures are filamentous. There is no firm conclusion about the magnetic field strength in galaxy filaments within the supercluster, neither from observations nor from theory \cite{2005xrrc.procE8.10D,Vazza:2015rqa,Vazza:2021vwy,Oei:2022kqx}. It seems that the field strength considered in \cite{Payez:2012vf} is overestimated, and the strict constraint would be relaxed to some extent. Despite this, it will help to restrict our results even if optical CP is produced in the supercluster.

\section{\label{sec:conclusions}Conclusions}
In this work, we investigate a potential optical CP in blazars induced by the ALP-photon mixing. Although this mixing is weak and has little effect on the flux and LP of photons, we have shown that it could induce a considerable optical CP when the oscillation length is much less than the coherent length of the magnetic field in the blazar jet. The formulas of the ALP induced CP in a simple conical jet are derived, which lead to a constraint on $\gag$, i.e., $\gag\cdot B_\mathrm{T0}\lesssim7.9\times10^{-12}~\mathrm{G\cdot GeV}^{-1}$ for $m_{a}\lesssim 10^{-13}~\mathrm{eV}$, corresponding to the upper limit on the observed CP at the level of $\sim0.1\%$. Hence, the coupling between the ALP and photon would be significantly constrained in the hadronic radiation model with a large magnetic field strength in the jet. Further research on BL Lacs and CP measurement in the optical band would significantly improve these results.

In particular, the tentative CP observations of two blazar 3C 66A and OJ 287 induced by ALP are analyzed using the data in the paper of Takalo and Sillanpaa. We find that these observations indicate a similar estimation of $\gag \sim 10^{-11}~\mathrm{GeV}^{-1}$. Large amount of simultaneous LP and CP measurements with high accuracy in the future are needed to verify the irregular results.

Though hadronic models set stringent constraint on $\gag$, the results could be significantly changed considering the intrinsic CP. Measuring the radio and optical CP simultaneously would be helpful to clarify the situation. Besides, note that many results of the ALP induced CP in this work are derived with an idealized jet configuration. As stated in the last part of Sec.\ref{sec:discussion}, we have shown that the obtained results are partially valid in more realistic cases.

\begin{acknowledgments}
 The authors would like to thank Zhao-Huan Yu, Zhuo Li, Kai Wang, Pu Du, and Jin Zhang for helpful discussions. The work is supported by the National Natural Science Foundation of China under grant No. 12175248. The work of JWW is supported by Natural Science Foundation of China under grant No. 12150610465 and the research grant “the Dark Universe: A Synergic Multi-messenger Approach” number 2017X7X85K under the program PRIN 2017 funded by the Ministero dell'Istruzione, Universit$\grave{a}$ e della Ricerca (MIUR).

\end{acknowledgments}

\hspace*{\fill}

\noindent\textit{Note added.}—During the completion of the paper, we became aware of a work \cite{Shakeri:2022usk} which also considers a similar topic. The axion-induced CP of light around black holes with some different physical settings is studied in that paper.
%bibliographystyle{IEEEtran}
\bibliography{OpticalCP}
\bibliographystyle{apsrev4-1}

\end{document}